\documentclass[longauth]{aa}  

\usepackage{graphicx}
\usepackage{upgreek}
\usepackage{natbib}
\usepackage{caption}
\usepackage{afterpage}
\usepackage{hyperref}
\usepackage{cleveref}
\usepackage[varg]{txfonts}
\begin{document}
   \title{Pulsar polarisation below 200 MHz: Average profiles and propagation effects}

   \titlerunning{Pulsar Polarisation Below 200 MHz}
   \authorrunning{A.~Noutsos et al.}
   
   \author{A. Noutsos\inst{\ref{mpifr}}
           \and C.~Sobey\inst{\ref{astron}}
           \and V.~I.~Kondratiev\inst{\ref{astron} \and \ref{lebedev}}
           \and P.~Weltevrede\inst{\ref{jbo}}
           \and J.~P.~W.~Verbiest\inst{\ref{bielefeld} \and \ref{mpifr}}
           \and A.~Karastergiou\inst{\ref{ox}}
           \and M.~Kramer\inst{\ref{mpifr} \and  \ref{jbo}}
           \and M.~Kuniyoshi\inst{\ref{mpifr}}
           \and A.~Alexov\inst{\ref{stsci}}
           \and R.~P.~Breton\inst{\ref{soton}}
           \and A.~V.~Bilous\inst{\ref{nijmegen}}
           \and S.~Cooper\inst{\ref{jbo}}
           \and H.~Falcke\inst{\ref{astron} \and \ref{nijmegen}}
           \and J.-M.~Grießmeier\inst{\ref{cnrs} \and \ref{nancay}}
           \and T.~E.~Hassall\inst{\ref{soton}}
           \and J.~W.~T.~Hessels\inst{\ref{astron} \and \ref{uva}}
           \and E.~F.~Keane\inst{\ref{swin} \and \ref{caastro}}
           \and S.~Os{\l}owski\inst{\ref{bielefeld} \and \ref{mpifr}}
           \and M.~Pilia\inst{\ref{astron}}
           \and M.~Serylak\inst{\ref{uwc}}
           \and B.~W.~Stappers\inst{\ref{jbo}}
           \and S.~ter Veen\inst{\ref{nijmegen}}
           \and J.~van Leeuwen\inst{\ref{astron} \and \ref{uva}}
           \and K.~Zagkouris\inst{\ref{ox}}
		\and K.~Anderson          \inst{\ref{aura}}
		\and L.~B\"ahren          \inst{\ref{nijmegen}}
		\and M.~Bell              \inst{\ref{fti} \and \ref{caastro}}
		\and J.~Broderick         \inst{\ref{ox}}
		\and D.~Carbone           \inst{\ref{uva}}
		\and Y.~Cendes            \inst{\ref{uva}}
		\and T.~Coenen            \inst{\ref{uva}}
		\and S.~Corbel            \inst{\ref{saclay}}
		\and J.~Eisl\"offel       \inst{\ref{tls}}
		\and R.~Fender            \inst{\ref{ox}}
		\and H.~Garsden           \inst{\ref{diderot}}
		\and P.~Jonker            \inst{\ref{sron} \and \ref{cfa} \and \ref{nijmegen}}
		\and C.~Law               \inst{\ref{berkley} \and \ref{uva}}
		\and S.~Markoff           \inst{\ref{uva}}
		\and J.~Masters           \inst{\ref{nrao} \and \ref{uva}}
		\and J.~Miller-Jones      \inst{\ref{curtin} \and \ref{uva}}
		\and G.~Molenaar          \inst{\ref{uva}}
		\and R.~Osten             \inst{\ref{stsci}}
		\and M.~Pietka            \inst{\ref{ox}}
		\and E.~Rol               \inst{\ref{uva}}
		\and A.~Rowlinson         \inst{\ref{cass}}
		\and B.~Scheers           \inst{\ref{uva} \and \ref{cwi}}
		\and H.~Spreeuw           \inst{\ref{uva}}
		\and T.~Staley            \inst{\ref{ox} \and \ref{soton}}
		\and A.~Stewart           \inst{\ref{ox}}
		\and J.~Swinbank          \inst{\ref{uva}}
		\and R.~Wijers            \inst{\ref{uva}}
		\and R.~Wijnands          \inst{\ref{uva}}
		\and M.~Wise              \inst{\ref{astron} \and \ref{uva}}
		\and P.~Zarka             \inst{\ref{meudon}}
		\and A.~van~der~Horst     \inst{\ref{uva}}
          }

   \institute{Max-Planck-Intitut f\"{u}r Radioastronomie, Auf dem H\"{u}gel 69, 53121 Bonn \\ \email{anoutsos@mpifr-bonn.mpg.de}\label{mpifr}
             \and
             ASTRON, the Netherlands Institute for Radio Astronomy, Postbus 2, 7990 AA Dwingeloo, The Netherlands\label{astron}
             \and
             Jodrell Bank Centre for Astrophysics, School of Physics and Astronomy, The University of Manchester, Manchester M13 9PL, UK\label{jbo}
             \and
             Fakult\"{a}t f\"{u}r Physik, Universit\"{a}t Bielefeld, Postfach 100131, 33501 Bielefeld, Germany\label{bielefeld}
             \and
             Oxford Astrophysics, Denys Wilkinson Building, Keble Road, Oxford OX1 3RH, UK\label{ox}
             \and
             Space Telescope Science Institute, 3700 San Martin Drive, Baltimore, MD 21218, USA\label{stsci}
             \and
             School of Physics and Astronomy, University of Southampton, SO17 1BJ, UK\label{soton}
             \and
             Department of Astrophysics/IMAPP, Radboud University Nijmegen, PO Box 9010, 6500 GL Nijmegen, The Netherlands\label{nijmegen}
             \and
             LPC2E - Université d'Orl\'{e}ans / CNRS\label{cnrs}
             \and
             Centre for Astrophysics and Supercomputing, Swinburne University of Technology, Mail H30, PO Box 218, VIC 3122, Australia\label{swin}
             \and
             Department of Physics \& Astronomy, University of the Western Cape, Private Bag X17, Bellville 7535, South Africa\label{uwc}
             \and
             Astro Space Centre, Lebedev Physical Institute, Russian Academy of Sciences, Profsoyuznaya Str. 84/32, Moscow 117997, Russia\label{lebedev}
             \and
             Station de Radioastronomie de Nan\c{c}ay, Observatoire de Paris, CNRS/INSU, USR 704 - Univ. Orléans, OSUC, 18330 Nan\c{c}ay, France\label{nancay}
             \and
             Anton Pannekoek Institute for Astronomy, University of Amsterdam, Science Park 904, 1098 XH Amsterdam, The Netherlands\label{uva}
             \and   
             ARC Centre of Excellence for All-sky Astrophysics (CAASTRO), 44 Rosehill Street, Redfern, NSW 2016, Australia\label{caastro}
            \and
            Gemini Observatory, Southern Operations Center, c/o AURA, Casilla 603, La Serena, Chile\label{aura}
            \and
            FTI Beverly, Massachusetts Office, Frontier Technology, Inc., 100 Cummings Ctr \#450G, Beverly, MA 01915, USA\label{fti}
            \and
            AIM-Unite Mixte de Recherche CEA-CNRS, Universit\`{e} Paris VII, CEA Saclay, Service d'Astrophysique, F-91191 Gif-sur-Yvette, France\label{saclay}
            \and
            Th\"{u}ringer Landessternwarte, Sternwarte 5, D-07778 Tautenburg, Germany\label{tls}
            \and
            Laboratoire AIM, Universit\`{e} Paris Diderot, Paris 7/CNRS/CEA-Saclay, DSM/IRFU/SAp, 91191 Gif-sur-Yvette, France\label{diderot}
            \and
            Harvard-Smithsonian Center for Astrophysics, 60 Garden Street, Cambridge, MA 02138, USA\label{cfa}
            \and
            Radio Astronomy Lab, UC Berkeley, CA, USA\label{berkley}
            \and
            National Radio Astronomy Observatory, 520 Edgemont Road, Charlottesville, VA 22903-2475, USA\label{nrao}
            \and
            International Centre for Radio Astronomy Research - Curtin University, GPO Box U1987, Perth, WA 6845, Australia\label{curtin}
            \and
            LESIA, Observatoire de Paris, CNRS, UPMC, Université Paris-Diderot, 5 place Jules Janssen, 92195 Meudon, France\label{meudon}
            \and
            CSIRO Astronomy and Space Science, PO Box 76, Epping, NSW 1710, Australia\label{cass}
            \and
            SRON, Netherlands Institute for Space Research, Sorbonnelaan 2, 3584 CA, Utrecht, the Netherlands\label{sron}
            \and
            Centrum Wiskunde \& Informatica, P.O.~Box 94079, 1090 GB Amsterdam, The Netherlands\label{cwi}
            }

   \date{Received October 20, 2014; accepted January 6, 2015}

 
  \abstract
   {}
   {We present the highest-quality polarisation profiles to date of 16 non-recycled pulsars and four millisecond pulsars, observed below 200 MHz with the LOFAR high-band antennas. Based on the observed profiles, we perform an initial investigation of expected observational effects resulting from the propagation of polarised emission in the pulsar magnetosphere and the interstellar medium.}
   {The polarisation data presented in this paper have been calibrated for the geometric-projection and beam-shape effects that distort the polarised information as detected with the LOFAR antennas. We have used RM Synthesis to determine the amount of Faraday rotation in the data at the time of the observations. The ionospheric contribution to the measured Faraday rotation was estimated using a model of the ionosphere. To study the propagation effects, we have compared our low-frequency polarisation observations with archival data at 240, 400, 600, and 1400 MHz.} 
   {The predictions of magnetospheric birefringence in pulsars have been tested using spectra of the pulse width and fractional polarisation from multifrequency data. The derived spectra offer only partial support for the expected effects of birefringence on the polarisation properties, with only about half of our sample being consistent with the model's predictions. It is noted that for some pulsars these measurements are contaminated by the effects of interstellar scattering. For a number of pulsars in our sample, we have observed significant variations in the amount of Faraday rotation as a function of pulse phase, which is possibly an artefact of scattering. These variations are typically two orders of magnitude smaller than that observed at 1400 MHz by Noutsos et al.~(2009), for a different sample of southern pulsars. In this paper we present a possible explanation for the difference in magnitude of this effect between the two frequencies, based on scattering.
Finally, we have estimated the magnetospheric emission heights of low-frequency radiation  from four pulsars, based on the phase lags between the flux-density and the PA profiles, and the theoretical framework of Blaskiewicz, Cordes \& Wasserman (1991). These estimates yielded heights of a few hundred km; at least for PSR B1133+16, this is consistent with emission heights derived based on radius-to-frequency mapping, but is up to a few times larger than the recent upper limit based on pulsar timing.
}
   {Our work has shown that models, like magnetospheric birefringence, cannot be the sole explanation for the complex polarisation behaviour of pulsars. On the other hand, we have reinforced the claim that interstellar scattering can introduce a rotation of the PA with frequency that is indistinguishable from Faraday rotation and also varies as a function of pulse phase. In one case, the derived emission heights appear to be consistent with the predictions of radius-to-frequency mapping at 150 MHz, although this interpretation is subject to a number of systematic uncertainties.}


   \maketitle
%

\section{Introduction}

The polarisation properties of pulsars below observing frequencies of a few hundred MHz are not very well known. A small number of publications have so far reported on properties below 300 MHz, such as the degree of fractional polarisation and the profiles of the polarisation position angle. In particular, Gould \& Lyne (1998)\nocite{gl98} published multi-wavelength polarisation profiles for 300 pulsars, between 230 and 1600 MHz, obtained with the Lovell telescope. However, polarisation information below 400 MHz was obtained for only $\approx 90$ pulsars in that work. More recently, Johnston et al.~(2008) used the Giant Meter-Wave (GMRT) and Parkes telescopes to study the polarisation of 67 bright pulsars between 243 and 3100 MHz. As in the preceding work, sensitivity and scattering limitations at low frequencies only allowed for 34 pulsars to be observable below 300 MHz. 

Below 200 MHz, pulsar polarisation observations have been sporadic and have focused on studies of individual, bright pulsars. Mainly, such observations have been performed with the Bol`shaya Steerable Array (BSA) of the Pushchino Radio Astronomy Observatory (PRAO; Shabanova \& Shitov 2004\nocite{ss04}; Suleymanova \& Rankin 2009\nocite{sr09}). However, the BSA is only sensitive to a single linear-polarisation sense and its frequency band is limited to $\approx 2$ MHz centred at 112 MHz. By combining BSA data with other low-frequency data, Shabanova \& Shitov~(2004) measured the interstellar Faraday rotation towards PSR B0950+08, while also estimating the ionospheric contribution to be less than 10\%. The total Faraday rotation was found to be ${\rm RM}=3$--6 rad m$^{-2}$, where RM is the rotation measure; this measure quantifies the strength of Faraday rotation towards a pulsar: it is the proportionality constant between the amount of rotation of the position angle of the linear polarisation (PA) and the observing wavelength squared, $\lambda^2$. Shabanova \& Shitov concluded that the published value by Taylor et al.~1993\nocite{tml93} (${\rm RM}=1.35\pm 0.15$ rad m$^{-2}$) is incorrect and that the actual RM value for this pulsar should be 3 times larger. Notably, both these measurements are consistent with the older and less constraining measurement by Hamilton \& Lyne (1987)\nocite{hl87}, being ${\rm RM}=2\pm 2$ rad m$^{-2}$. The most recent measurement by Johnston et al.~(2005)\nocite{jhv+05}, using polarisation data from Parkes, at 1400 MHz, yielded the much more constraining value of ${\rm RM}=-0.66\pm 0.04$ rad m$^{-2}$. It should be stressed that the published values by Johnston et al.~(2005) and by Hamilton \& Lyne (1987) were corrected for the ionospheric Faraday rotation, ${\rm RM}_{\rm iono}$, by means of subtracting it from the total RM. In the work of Johnston et al., ${\rm RM}_{\rm iono}$ was estimated to be $-2$--0 rad m$^{-2}$. It is not clear whether this correction was also made by Taylor et al.~(1993).

The significant differences in the RM value of PSR B0950+08 from the different observations --- at least where the ionospheric contribution was taken into account --- is not clear. Nevertheless, such inconsistencies between the RM values of individual pulsars have also been reported in more recent studies (Noutsos et al.~2008\nocite{njkk08}). The most recent measurements by Shabanova \& Shitov and Johnston et al.~accounted for the ionospheric contribution and were nearly contemporaneous, which makes it unlikely that the RM difference is due to changes in the interstellar medium (ISM) or due to the pulsar's relative motion to the observer. Changes in the local interplanetary medium or even calibration errors could be the reasons behind those differences, although the former would most likely be responsible for only a small fraction of a rad m$^{-2}$ for pulsar observations several solar radii away from the Sun (You et al.~2012\nocite{ych+12}).

According to a simple picture of magnetospheric production of radio emission, the wavelength of pulsar radio emission is related to the local plasma density in the open field-line region (Ruderman \& Sutherland 1975\nocite{rs75}). This is the so-called {\em radius-to-frequency mapping (RFM)} model, which implies that high-frequency emission is generated closer to the polar caps than low-frequency emission and, thus, the observed spectrum of radio frequencies traces a range of emission altitudes above the polar caps. Based on pulsar timing and polarisation measurements, it has been estimated that for the majority of pulsars for which this has been measured the range of altitudes across which pulsar radio emission is generated ranges from a fraction of a percent to a few percent of the light-cylinder radius, $R_{\rm LC}=cP/(2\pi)$ --- where $P$ is the pulsar spin period and $c$, the speed of light (Cordes 1978\nocite{cor78}; Weltevrede \& Johnston~2008\nocite{wj08}; Hassall et al.~2012\nocite{hsh+12}). It should be noted that for a few pulsars this value has been found to be up to several tens of percents. 

Furthermore, a number of studies suggest that pulsars exhibit lower fractions of linear polarisation towards high frequencies. An early study of the linear polarisation of 20 pulsars by Manchester et al.~(1973)\nocite{mth73}, between $\approx 100$ MHz and a few GHz, suggested that pulsars are more highly polarised at low frequencies. Interestingly, for several pulsars it was observed that the polarisation fraction is roughly constant up to a critical frequency, above which the polarisation decreases linearly as a function of observing frequency. Later, Xilouris et al.~(1996)\nocite{xkj+96} investigated the spectrum of linear-polarisation fractions for 8 bright pulsars, between $\approx 100$ MHz and 32 GHz. In most cases, it was shown that pulsars depolarise rapidly towards high frequencies, while nearly half the sample of pulsars investigated also exhibited a spectral steepening of the degree of depolarisation towards the highest frequencies. Such studies motivated an explanation for the frequency-dependent depolarisation of pulsars, an attempt for which was provided by von Hoensbroech, Lesch \& Kunzl 1998\nocite{hlk98}, who interpreted the phenomenon in terms of the birefringence of the magnetospheric plasma (see Section~\ref{sec:freqevo}). Notably, the role of birefringence in pulsar magnetospheres had been suggested much earlier, in the work of Novick et al.~(1977)\nocite{nwas77}. Johnston et al.~(2008)\nocite{jkm+08} note that depolarisation at high frequencies may simply be related to the fact that high-frequency radio emission traverses longer paths through the magnetosphere, in the framework of RFM. 
 
Nevertheless, the work of Gould \& Lyne~(1998) and Johnston et al.~(2008) found several cases where a simple, monotonic relationship between the polarisation fraction and the observing frequency is not followed. Moreover, in certain cases, the degree of polarisation remains roughly constant throughout the explored frequency range. The above authors put forward geometrical arguments to explain the depolarisation at higher frequencies, arguing that several, short but highly polarised bursts of emission are incoherently summed at the detector, during the sampling interval. At higher frequencies, the emission is generated at lower altitudes where the magnetic-field density is higher, leading to a higher number of incoherently summed bursts whose average polarisation is lower. Alternatively, it has been suggested that the superposition of orthogonal modes of linearly polarised emission with different spectral indices can also cause depolarisation at higher frequencies (Karastergiou et al.~2005\nocite{kjm05}).

The work of Johnston et al.~(2008) used a sample of 34 pulsars that exhibited low scattering at 243 MHz (quantified by the pulse broadening between 3.1 GHz and 243 MHz). However, the effects of scattering have been seen in polarisation at 1400 MHz, even in cases where the total power appears little or moderately scattered (Karastergiou 2009\nocite{kar09}; Noutsos et al.~2009\nocite{nkk+09}). Karastergiou showed through simulations that scattering causes flattening of steep gradients in PA profiles and, furthermore, that scattering is a plausible explanation for the observed variations in the amount of Faraday rotation as a function of pulse phase, which had been observed by Noutsos et al.~(2009). The main reason for these effects is the different degrees of superposition of linearly polarised intensity between earlier and later pulse phases (due to scattering), as a function of frequency. The magnitude of these effects is expected to increase dramatically with decreasing frequency, as scattering scales proportionally to $f^{-4}$ (Cronyn 1970; but also see \nocite{bcc+04}Bhat et al.~2004). However, the frequency evolution of some of these effects, e.g.~phase-resolved RM variations due to scattering, has not yet been investigated (see Section~\ref{sec:rmvars}).

The following sections contain the presentation and analysis of the first data set of 20 polarised pulsars observed with the high-band antennas (HBAs) of the Low Frequency Array (LOFAR). The content of this paper is set out in the following way. Section~\ref{sec:obs} describes our observing set-up and the calibration procedure. In Section~\ref{sec:datcal}, we provide the methods that were used to test the quality of the polarisation calibration of pulsar data obtained with LOFAR. The data set of 20 polarisation profiles is presented in Section~\ref{sec:polprofs}. The data analysis performed in this paper is concerned with the investigation of two propagation effects that may affect linearly polarised radio emission between the pulsar and the telescope: (a) Section~\ref{sec:freqevo}: the effects of the birefringence of magnetospheric plasma on the frequency evolution of pulsar polarisation profiles; (b) Section~\ref{sec:rmvars}: the effects of interstellar scattering on the observed linearly polarised emission and the measured Faraday rotation. In the Discussion, Section~\ref{sec:emheights}, we provide estimates of the emission altitudes of radio emission, based on the phase lag between the observed emission and the location of the emission in the co-rotating magnetosphere; we also individually discuss the polarisation properties of three of the four millisecond pulsars (MSPs) that we observed with LOFAR. Finally, in Section~\ref{sec:summary} we summarise this paper and draw conclusions based on the results of our analysis.

\section{Observations}
\label{sec:obs}
LOFAR is an international interferometric telescope, composed of many thousands of dipole antennas grouped into stations. Each station comprises two types of LOFAR antennas, the low-band antennas (LBA) and high-band antennas (HBA), which are sensitive to 10--90 MHz and 105--240 MHz radio frequencies, respectively. The LOFAR stations are arranged in a sparse array, spread across Europe, with a dense core region located in the Netherlands; at the centre of this core region there is an isolated complex of six LOFAR stations, called the {\em Superterp}. For a general LOFAR description see van Haarlem et al.~(2013)\nocite{vwg+13}, and for a full description of how LOFAR is used for pulsar observations see Stappers et al.~(2011)\nocite{sha+11}. 

Our pulsar polarisation observations were performed in November and December 2012, with 24 stations of the LOFAR core. We observed 20 bright pulsars, which were selected (a) based on their high flux densities at 102.5 MHz, published by Malofeev, Malov \& Shchegoleva~(2000)\nocite{mms00} who performed observations with the Large Phased Array of the Lebedev Institute of Physics; (b) based on their high degrees of linear polarisation, as derived from polarisation observations between 230 and 1600 MHz by Gould \& Lyne (1998)\nocite{gl98}; and (c) based on their relatively high source declination, which ensured that the pulsars could be observed at an elevation of $>30^\circ$, minimising the complexity of correcting for the elevation-dependent effects of LOFAR's sensitivity (see Section~\ref{subsubsec:calperf}). The list of 20 pulsars observed at 150 MHz for this paper is shown in the first column of Table~\ref{tab:polproperties}.

The typical integration time per pulsar in our observations was 10 minutes. Our observing set-up used $\Delta f=92$ MHz of instantaneous bandwidth between $f_{\rm min}=105$ and $f_{\rm max}=197$ MHz, centred at $f_{\rm c}=150$ MHz and split into 470$\times$195 kHz subbands. Each subband contained the raw signal sampled as complex X- and Y-sense voltages, at the baseband temporal resolution of 5.12 $\upmu$s. The large available bandwidth was the result of recording the voltage data as 8-bit samples, instead of the standard 16 bits/sample, which would result in half the above bandwidth, given hardware limitations on the total data rate. For these observing parameters, the minimum detectable flux of the LOFAR core at 150 MHz is $\approx 0.5$ mJy. The details of the online processing for pulsar observations with LOFAR are described in Stappers et al.~(2011)\nocite{sha+11}.

\begin{table*}
\caption{The polarisation properties of the 20 pulsars presented in this paper, as measured from LOFAR observations, between 105 and 197 MHz. The duration of each observation is shown in Col.~2. The 1$\sigma$ statistical uncertainties, shown in parentheses, refer to the last significant digit of the tabulated values. Column 3 shows the published value of the RM, from the ATNF pulsar catalogue. Column 4 shows the value of the RM obtained from LOFAR data, using RM Synthesis. The quoted uncertainties are purely statistical and do not incorporate systematics due to e.g.~the ionospheric and solar-wind contributions. Column 5 shows the contribution of the ionosphere to the measured RM values from LOFAR, calculated for each observation, from the model of Sotomayor-Beltran et al.~(2013). The last three columns show the linear, circular and total polarisation fraction in the average profiles from LOFAR. The quoted uncertainties are purely statistical and do not incorporate a systematic uncertainty of 5--10\%, attributable to the calibration model (see Section~\ref{subsubsec:calperf2}). We note that because of the lack of significant linear polarisation, it was not possible to meaure an RM for PSRs J0034$-$0534 and B2111+46. The corresponding linear-polarisation fraction for PSR J0034$-$0534 was calculated assuming the ${\rm RM}=0$ rad m$^{-2}$, which only yields an upper limit due to significant instrumental polarisation (see Section~\ref{subsubsec:mspj0034}); the linear-polarisation fraction for PSR B2111+46 was calculated assuming the published RM.} 
\label{tab:polproperties}      
\centering                          
\begin{tabular}{l r r r r r r r}        
\hline\hline                 
PSR & Duration [min] & RM$_{\rm pub}$ [rad m$^{-2}$] & RM$_{\rm LOFAR}$ [rad m$^{-2}$] &  RM$_{\rm iono}$ [rad m$^{-2}$] & $L$[\%] & |$V$|[\%] & $L$+|$V$|[\%]  \\    
\hline                        
B0031$-$07   & 10       & 9.8(2)$^{1}$      & 10.977(4)    & 1.09(7)     & 41.1(1)          & 9.8(2)   & 50.0(3)  \\
J0034$-$0534 & 20       & --                & --                & 1.10(7)     & $<$15.7(2)$^{*}$ & 12.0(2)  & $<$27.7(3)$^{*}$  \\
B0136+57     & 10       & $-$90(4)$^{1}$    & $-$93.689(6) & 0.44(8)     & 47.3(3)          & 8.7(5)   & 56.0(6)  \\
B0809+74     & 10       & $-$11.7(13)$^{2}$ & $-$13.566(1) & 0.43(7)     & 18.43(1)         & 4.18(3)  & 22.61(3) \\
B0823+26     & 10       & 5.9(3)$^{3}$      & 5.942(3)     & 0.56(6)     & 25.18(4)         & 6.21(7)  & 31.4(1)  \\
B0834+06     & 10       & 23.6(7)$^{1}$     & 26.095(1)    & 0.77(7)     & 25.45(2)         & 2.59(4)  & 28.05(5) \\
B0950+08     & 10       & $-$0.66(4)$^{4}$  & 2.151(1)     & 0.67(6)     & 73.9(1)          & 11.54(5) & 85.4(1)  \\
J1012+5307   & 20       & --                & 3.38(1)      & 0.40(6)     & 92(1)            & 9(1)     & 100(2)    \\
J1022+1001   & 20       & $-$0.6(5)$^{5}$   & 2.18(2)      & 0.79(5)     & 82(2)            & 14(2)    & 96(2)    \\
B1133+16     & 10       & 1.1(2)$^{4}$      & 4.770(1)     & 0.80(7)     & 33.92(1)         & 17.00(4) & 50.93(4) \\
B1237+25     & 10       & -0.33(6)$^{6}$    & 0.864(2)     & 1.06(6)     & 46.3(1)          & 7.5(1)   & 53.8(2)  \\
B1257+12     & 30       & --                & 9.24(3)      & 1.33(6)     & 24.2(4)          & 18(1)    & 42(1)    \\
B1508+55     & 10       & 0.8(7)$^{2}$      & 2.449(2)     & 1.17(6)     & 10.21(1)         & 6.57(3)  & 16.78(4)  \\
B1911$-$04   & 10       & 4.4(9)$^{4}$      & 6.23(2)      & 2.25(5)     & 16.1(2)          & 7.0(4)   & 23.1(5)  \\
B1919+21     & 10       & $-$16.5(5)$^{1}$  & $-$16.104(2) & 0.89(5)     & 18.74(3)         & 6.82(7)  & 25.6(1)  \\
B1929+10     & 10       & $-$6.87(2)$^{4}$  & $-$5.841(2)  & 1.11(4)     & 87.0(3)          & 22.8(3)  & 109.9(4) \\
B1953+50     & 10       & $-$22(2)$^{1}$    & $-$23.07(1)  & 0.77(5)     & 19.6(2)          & 6.0(5)   & 25.6(5)  \\
B2111+46     & 10       & $-$224(2)$^{2}$   & --           & 0.53(6)     & 1.423(4)$^{**}$  & 5.6(2)   & 7.1(2)$^{**}$ \\
B2217+47     & 10       & $-$35.3(18)$^{2}$ & $-$35.407(2) & 0.52(6)     & 19.33(2)         & 9.17(5)  & 28.5(1)  \\
B2224+65     & 10       & $-$21(3)$^{1}$    & $-$22.486(8) & 0.50(7)     & 48.1(4)          & 8.7(8)   & 56.7(9)    \\   
\hline                                   
\end{tabular}
\\
\flushleft Published RM references: 1.\,Hamilton \& Lyne (1987)\nocite{hl87} \ 2.\,Manchester (1972)\nocite{man72} \ 3.\,Manchester (1972)\nocite{man74} \ 4.\,Johnston et al.~(2005)\nocite{jhv+05} \ 5.\,Yan et al.~(2011)\nocite{ymv+11} \ 6.\,Taylor et al.~(1993)\nocite{tml93}
\flushleft *calculated at ${\rm RM}=0$ rad m$^{-2}$ \\
\flushleft **calculated at the published RM \\
\end{table*}

\section{Data reduction and polarisation calibration}
\label{sec:datcal}

\subsection{Data reduction}
The complex data within each 195 kHz subband were coherently de-dispersed (Hankins \& Rickett 1975\nocite{hr75}) using the pulsar's known dispersion measure (DM), published in the Australia Telescope National Facility (ATNF) pulsar catalogue\footnote{http://www.atnf.csiro.au/people/pulsar/psrcat/}, in order to correct the effects of interstellar dispersion. The coherent de-dispersion was performed using the DSPSR digital signal-processing software (van Straten \& Bailes 2011\nocite{vb11}). The de-dispersed signal was folded with the known timing ephemeris obtained from the ATNF pulsar catalogue and the data were further reduced by averaging the signal over every 5 seconds of data (i.e.~a subintegration). The down-sampled data were transformed from XY auto- and cross-correlations to Stokes I,Q,U,V parameters and written out as a PSRFITS\footnote{http://www.atnf.csiro.au/research/pulsar/psrfits/} archive. Finally, impulsive radio frequency interference (RFI) from sources of terrestrial origin was excised by visually determining and zero-weighting the affected subbands and subintegrations. On average, no more than 5\% of each data set was zero-weighted during this step. For some of the pulsars, it was necessary to fine-tune the value of the DM by maximising the S/N of the average pulse profile over a small range of values around the published DM. All of the above steps were performed using the PSRCHIVE pulsar processing suite (Hotan et al.~2004\nocite{hvm04}).  

\subsection{Instrumental calibration}
The sensitivity of LOFAR decreases significantly away from the zenith, mainly due to the signal projection onto the ground-fixed antennas. In addition, the recorded pulsar polarisation is prone to geometric distortions caused by parallactic rotation between the polarisation plane of the pulsar emission and the fixed orientation of the antenna dipoles, due to the Earth's diurnal rotation. In addition, the polarisation of the signal is affected by the difference in sensitivity between the X and Y senses of the antennas. These instrumental effects need to be removed, in order to study the intrinsic polarisation properties of the observed pulsars. 

The current model that describes the sensitivity of LOFAR as a function of direction and observing frequency is based on electromagnetic (EM) simulations of the antenna gains, using as its basis the measurement equations by Hamaker et al.~(1996)\nocite{hbs96} (see e.g.~Smirnov 2011\nocite{smi11}a). The calculation of the model's parameters for any frequency and spatial direction is done by polynomial fits and Taylor expansions of the simulations, respectively. We note that the current model contains only the beam and gain corrections of the full Jones formalism: in the Jones calculus, these are expressed with the \textsf{\textbf{E}} and \textsf{\textbf{G}} complex matrices (Jones 1941\nocite{jon41}). The pulsar profiles were calibrated by applying the inverse of the instrumental response, as is expressed by the Jones matrix for each subband and each subintegration, to the Stokes profiles of the pulsar. 

\subsubsection{Calibration performance: Sensitivity}
\label{subsubsec:calperf}
\begin{figure*}
\centering
\includegraphics[height=\dimexpr \textheight - 10\baselineskip\relax,width=\textwidth,keepaspectratio]{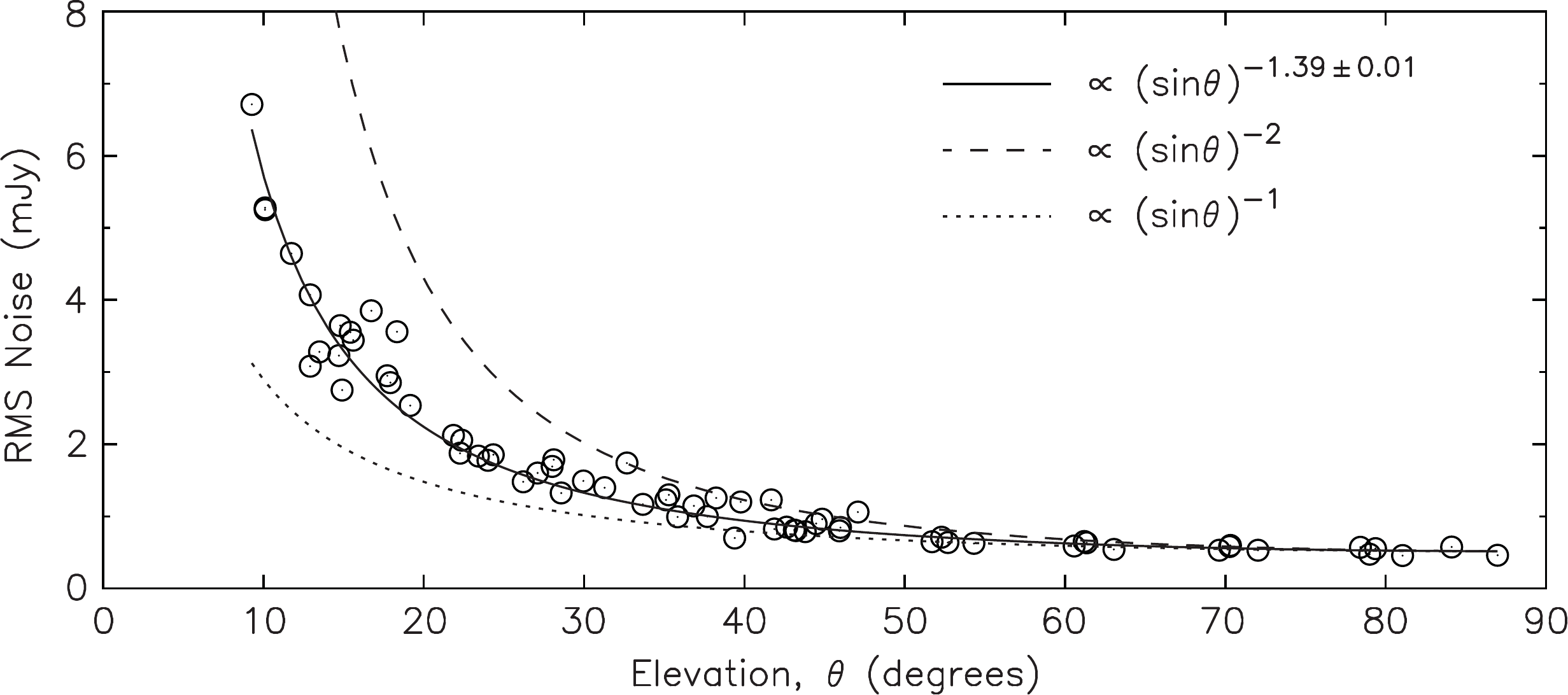}
\caption{Post-calibration dependence of the sensitivity of LOFAR on source elevation, measured during the observation of 4 pulsars (PSRs B0834+06, B1929+10, B1953+50 and B2217+47) at various hour angles, with the LOFAR core. Here, the sensitivity is expressed as the RMS of the off-pulse flux density, which has been normalised to $0.5$ mJy at elevation of $\theta=90^\circ$; this flux density corresponds to the theoretical sensitivity of the instrument, assuming the parameters of our observations (see text). The best fit of a power law on $1/\sin\theta$ to the data, as is expected by a simple signal projection, is shown with a solid curve. For comparison, we show the $1/\sin\theta$ function (dotted curve), normalised in the same way, which is the expected theoretical dependence from Lambert's law of a simple signal projection of unpolarised emission. The expected theoretical dependence of fully polarised emission, assuming an ideal dipole antenna follows Malus's law, $1/\sin^2\theta$ (dashed curve).}
\label{fig:elevsens}%
\end{figure*}

We have measured the model's performance as a function of source elevation, with observations of 4 bright pulsars, PSRs B0834+06, B1929+10, B1953+50 and B2217+47, at elevations between $\approx 9^\circ$ and $87^\circ$. These auxiliary observations used the same observing set-up as that for the 20 pulsars of our main sample. After performing the Hamaker calibration on each data set, we generated time- and frequency-averaged pulse profiles for each observation. We then measured the RMS of the off-pulse noise in each pulsar's averaged flux-density profile. In order to account for the differences in the amount of integration and bandwidth between observations, due to RFI excision, we used the radiometer equation to scale the RMS values from all observations to the same frequency bandwidth and integration time. Finally, we plotted the RMS values as a function of elevation, $\theta$ (Fig.~\ref{fig:elevsens}). As we have not performed absolute flux calibration, we normalised the flux scale to be 0.5 mJy at $\theta=90^\circ$, corresponding to the expected direction-independent sensitivity of our observing set-up. Our measurements show that below a source elevation of 20$^\circ$ the sensitivity decreases by at least a factor of $\approx 4$. However, above 35$^\circ$ the sensitivity remains within a factor of 2 of its value at zenith. The observations of the 20 pulsars presented in this paper took place near the transit time of each pulsar. As a result, all pulsars were above 30$^\circ$ in elevation, during each observation.

Ideally, if the calibration model perfectly describes the LOFAR antennas, after calibration the RMS noise should be independent of elevation. Our tests have shown that this remains true for $\theta>45^\circ$, to within 18\% accuracy. However, below that limit the model deviates significantly from a flat response. In Fig.~\ref{fig:elevsens}, we have drawn the expected dependence of sensitivity on elevation of a ground-fixed antenna, in two cases: (a) the case of a simple projection of an unpolarised signal, given by Lambert's cosine law ($\propto 1/\sin\theta$); (b) the case of a projection of a 100\% linearly polarised signal onto a co-polarised dipole antenna, given by Malus's law ($\propto 1/\sin^2\theta$). It can be seen that although calibration cannot recover the full sensitivity at low elevations, the reduction of sensitivity with decreasing elevation is much less than for an uncalibrated dipole: a power-law fit on $\sin\theta$ to the data yields that the RMS noise scales as $1/\sin^{1.39}\theta$. Even at elevations below $20^\circ$, the post-calibration sensitivity is $>50\%$ better than without calibration.

\begin{figure}
\centering
\includegraphics[height=\dimexpr \textheight - 15\baselineskip\relax,width=0.47\textwidth,keepaspectratio]{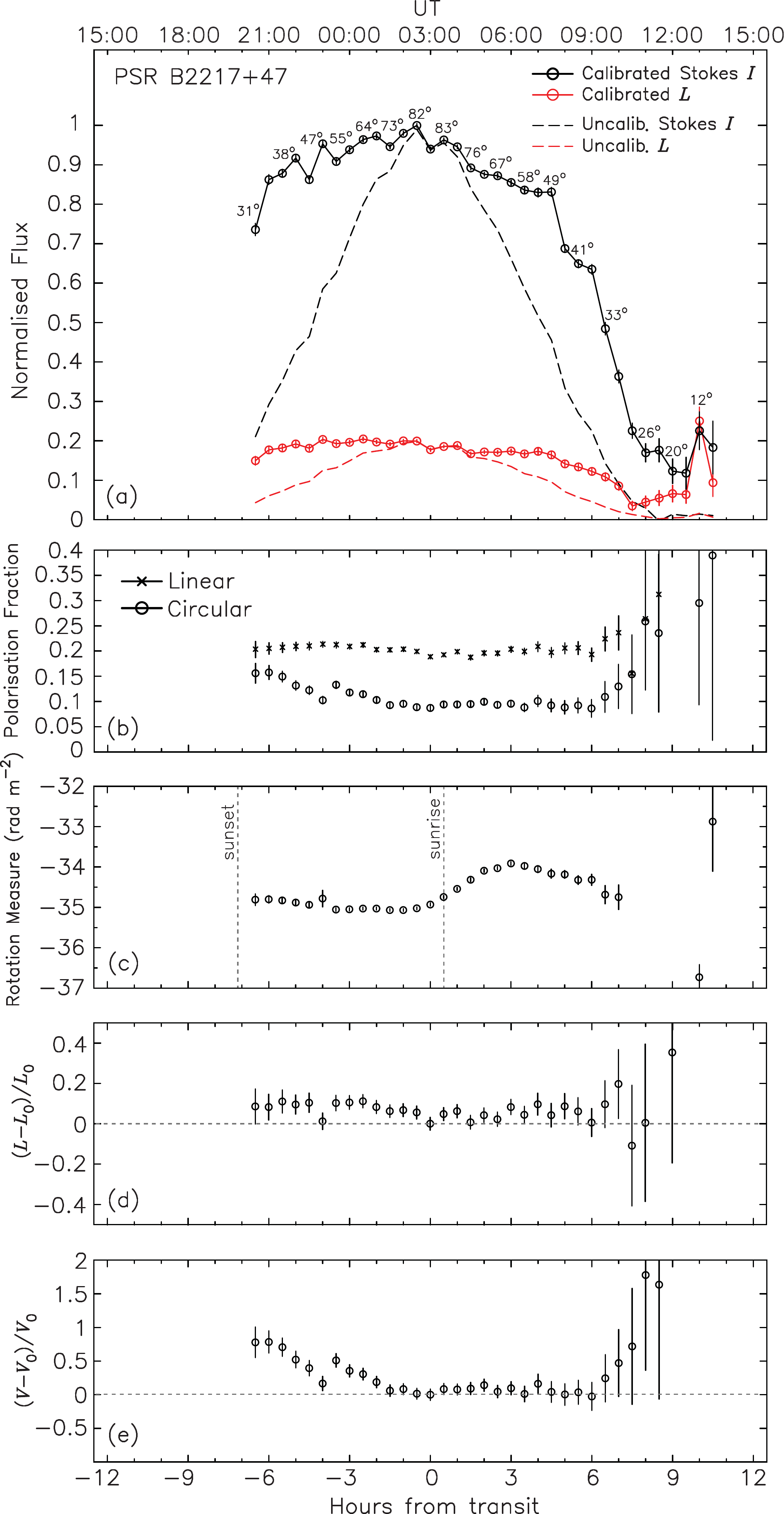}
\caption{The total and polarised flux density of PSR B2217+47 as a function of hour angle, mapped during the 17-hour test observation to evaluate the performance of the calibration model for LOFAR. Each data point corresponds to a 15-minute integration, every 30 minutes. Source transit occurred at 03:00 UT, corresponding to 0 hours from transit on the lower $x$-axis. Panel (a) shows the pulse-averaged flux densities before (dashed black line) and after beam calibration (black circles). Also shown are the pulse-averaged linearly polarised flux densities before (dashed red lines) and after beam calibration and correction for Faraday rotation (red circles), using the corresponding RM values shown in plot (c). All values in panel (a) have been normalised by the maximum flux value across the observation and each point has been labelled with the elevation of the pulsar at the beginning of each observation. Panel (b) shows the linear (crosses) and circular (circles) polarisation fractions after calibration and Faraday rotation correction. Panel (c) shows the observed RM varying due to ionospheric Faraday rotation over the 17-hour timespan. Times of sunset and sunrise are also marked with grey dashed lines. Panels (d) and (e) show the fractional change in the pulse-averaged linear polarisation and the pulse-averaged circular polarisation, respectively.}
\label{fig:B2217beamtests}%
\end{figure}

\begin{figure}
\centering
\includegraphics[height=\dimexpr \textheight - 10\baselineskip\relax,width=0.47\textwidth,keepaspectratio]{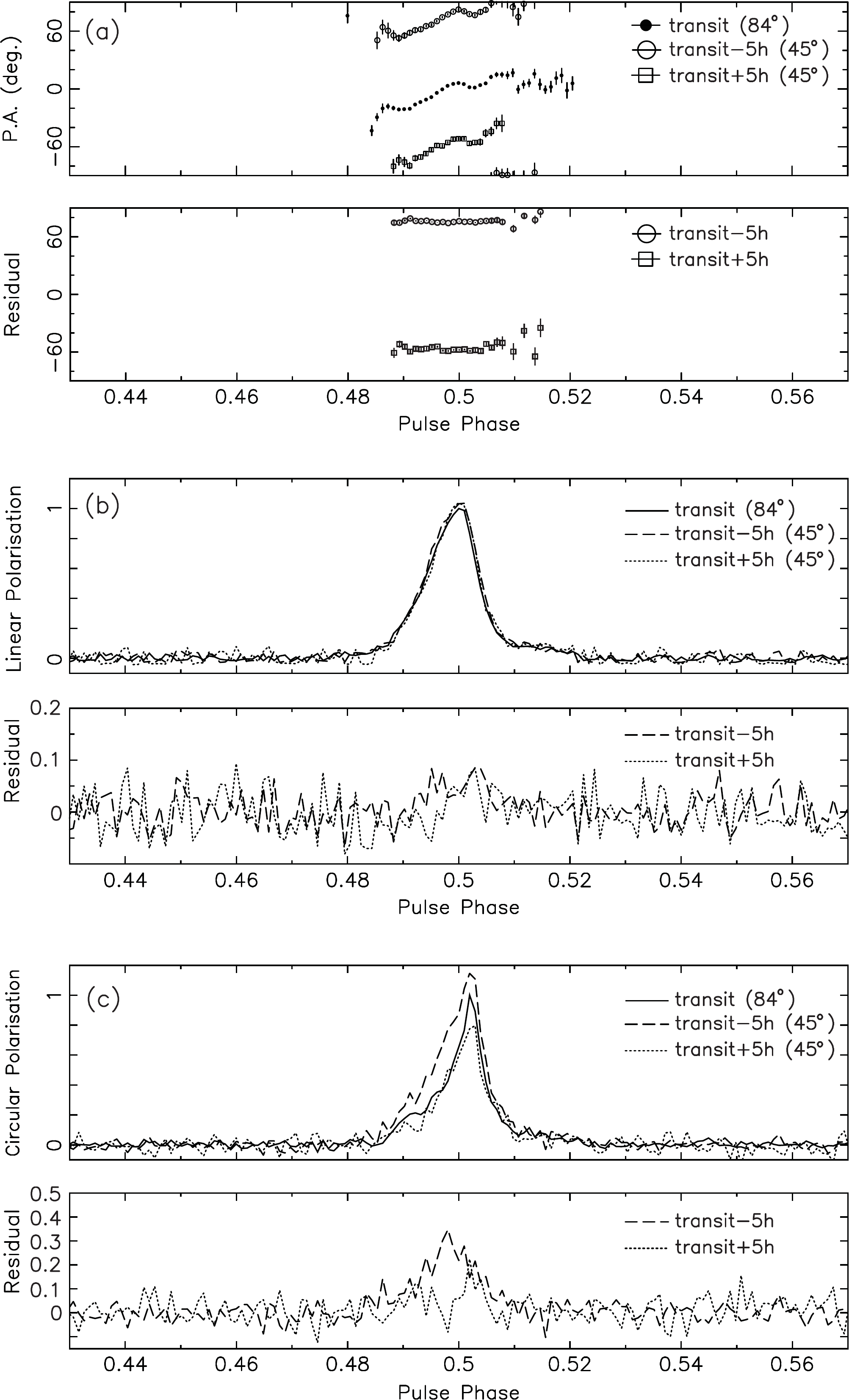}
\caption{Comparison of the polarisation profiles of PSR B2217+47 between 3 observations: when the pulsar was at transit and 5 hours before and after transit. The transit observation corresponds to $\approx 84^\circ$ elevation and those around transit, to $\approx 45^\circ$. (a) Comparison between the polarisation-angle profiles, where the phase-independent offsets between the profiles are due to the parallactic rotation of the source with respect to the LOFAR antennas. (b),(c) Comparison between the linear and circular polarisation profiles, where the flux density has been normalised by the maximum flux value of the profile at transit. Below the profiles, the residual difference between the profiles around transit and that at transit is shown.}
\label{fig:B2217profdiffs}%
\end{figure}

\subsubsection{Calibration performance: Polarisation}
\label{subsubsec:calperf2}
As was mentioned in the previous section, the sensitivity of LOFAR depends strongly on observing direction. Furthermore, since a large fraction of pulsar emission is polarised, the amount of polarisation projected onto each antenna dipole can vary significantly as the pulsar moves across the sky. Hence, before we can draw conclusions about the degree of polarisation of the observed pulsars, we must correct for the direction-dependent gains of the LOFAR antennas. The performance of the beam model, which we used to correct the data, was tested with a 17-hour observation of the bright pulsar, PSR B2217+47, with the Effelsberg HBA station (codenamed DE601). The choice of observatory was based on the availability of DE601 for long test observations, being higher than for the LOFAR core. During the observation, we recorded $\delta t=15$ minutes of data, every 30 minutes. On average, each integration corresponded to approximately 1,700 pulsar rotations, thus yielding stable profiles that should be relatively unaffected by pulse-to-pulse intensity variations. The available bandwidth during these test observations was approximately $\delta f=36$ MHz, ranging from 127 to 163 MHz. Our observations sampled elevations ranging from 12$^\circ$, which roughly corresponds to the south-west horizon defined by the surrounding hills of the Effelsberg site, up to 84$^\circ$, near transit. It should be noted that the DE601 horizon at the south-east location where this pulsar rises corresponds to an elevation of $\approx 30^\circ$. In that respect, the topology of DE601 is unique and unlike the Dutch stations that were used for our main observations: the latter are unobstructed by the surrounding terrain and allow for observations at elevation angles that are lower than those accessible with DE601, in all azimuthal directions. For each 15-minute pointing, we calculated the average values of the total and linearly polarised flux density across the pulse. These were calculated as
\begin{align}
\label{eq:ILstats1}
\langle I \rangle &= \frac{1}{N_{\rm ON}}\sum_{\rm ON}I \\
\langle L \rangle &= \frac{1}{N_{\rm ON}}\sum_{\rm ON}L
\end{align}
where, according to Everett \& Weisberg (2001)\nocite{ew01},
\begin{align}
L &= \sigma_I\sqrt{\frac{Q^2+U^2}{\sigma_I^2}-1} \ \ \ \ \ \ \ {\rm if} \  \frac{\sqrt{Q^2+U^2}}{\sigma_I}>1.57 \\ 
L &= 0 \ \ \ \ \ \ \ \ \ \ \ \ \ \ \ \ \ \ \ \ \ \ \ \ \ \ \ \ \ \ \ \ \ \ {\rm otherwise,}
\end{align}
where $\sigma_I$ is the off-pulse RMS of Stokes $I$, calculated across $N_{\rm OFF}$ phase bins and scaled to the width of the on-pulse area; $\sum_{\rm ON}$ denotes bin-wise summation of the respective quantity across the on-pulse area, corresponding to $N_{\rm ON}$ bins.

In addition, we calculated the uncertainties on the above quantities as   
\begin{align}
\label{eq:ILstats2}
\sigma_{\langle I \rangle} &= \frac{1}{\sqrt{N_{\rm ON}}}\sqrt{\sum_{\rm OFF}\frac{I^2}{N_{\rm OFF}}} \\
\sigma_{\langle L \rangle} &= \frac{1}{\sqrt{N_{\rm ON}}}\sqrt{\sum_{\rm ON}\left(\frac{Q}{L}\right)^2\sigma_Q^2+\sum_{\rm ON}\left(\frac{U}{L}\right)^2\sigma_U^2},
\end{align}
where $\sigma_Q$ and $\sigma_U$ are the off-pulse RMS values of Stokes $Q$ and $U$, respectively, scaled to the width of the on-pulse area.

The average on-pulse flux and linear polarisation were calculated before and after applying the beam model corrections. In order to account for the different amounts of RFI excision to which the data were subjected before the flux calculations, we scaled all fluxes by the effective bandwidth and integration time of each observation, i.e.~according to the radiometer equation, $\langle I\rangle \propto (\delta f_{\rm eff} \cdot \delta t_{\rm eff})^{-1/2}$. Figure~\ref{fig:B2217beamtests}a shows the calibrated and uncalibrated flux for each pointing, as a function of hour angle relative to the time of transit (corresponding to 0). It should be noted that each observation was corrected for the effect of Faraday rotation before calculating the amount of linear polarisation. This was done by calculating an RM separately for each pointing, as the ionospheric contribution could vary significantly over several hours (Sotomayor-Beltran et al.~2013\nocite{ssh+13}). Figure~\ref{fig:B2217beamtests}c shows the RM values with which we corrected the data, as a function of hour angle. The variability of the ionospheric contribution to the measured Faraday rotation is evident in that plot: it can be seen that ${\rm RM}_{\rm iono}$ varies by $\approx 1$ rad m$^{-2}$ during the 17-hour observation, with the maximum occurring roughly 3 hours after sunrise. Furthermore, in order to check the amount of leakage between total intensity and linear polarisation at different elevations, we calculated the fraction of linear polarisation at each pointing: $\langle L\rangle/\langle I\rangle$. The uncertainty on $\langle L\rangle/\langle I\rangle$ can be expressed as
\begin{equation}
\label{eq:polfracerr}
\sigma_{L/I}=\frac{\langle L\rangle}{\langle I\rangle}\sqrt{\left(\frac{\sigma_{\langle L \rangle}}{\langle L \rangle}\right)^2+\left(\frac{\sigma_{\langle I \rangle}}{\langle I \rangle}\right)^2}.
\end{equation}
The polarisation fraction as a function of hour angle, for our observation, is shown in Fig.~\ref{fig:B2217beamtests}b. It can be seen that the fraction of linear polarisation remains constant across the entire observation, with a mean value of 20.3(7)\%.
On the other hand, the fraction of circular polarisation shown in the same figure is less stable across the entire observation but is well-behaved for the majority of pointings. More specifically, including all pointings within 3 hours either side of transit, we obtain a mean circular-polarisation fraction of 9.7(2)\%.

Another interesting diagnostic is the fractional difference of the linear and circular polarisation between its value at transit, $L_0$ and $V_0$, and those at lower elevations: i.e.~$(L-L_0)/L_0$ and $(V-V_0)/V_0$, respectively. Ideally, the difference should be consistent with 0. However, imperfect calibration would lead to Stokes $I$ leaking power into $L$ and $V$. Indeed, the diagnostic plot in Fig.~\ref{fig:B2217beamtests}d shows that for pointings with elevation $>30^\circ$, the mean fractional difference of $L$ is 6.0(7)\%. The impact of imperfect calibration is higher for the circular polarisation. The results show a steep increase in the amount of $V$, for eastwards pointings up to 2 hours before transit (elevations $<70^\circ$), but remain roughly constant with a mean of 7(2)\%, for the rest of the observation. This asymmetry in the performance of the model is not well understood. 

\begin{figure}
\centering
\includegraphics[height=\dimexpr \textheight - 10\baselineskip\relax,width=0.47\textwidth,keepaspectratio]{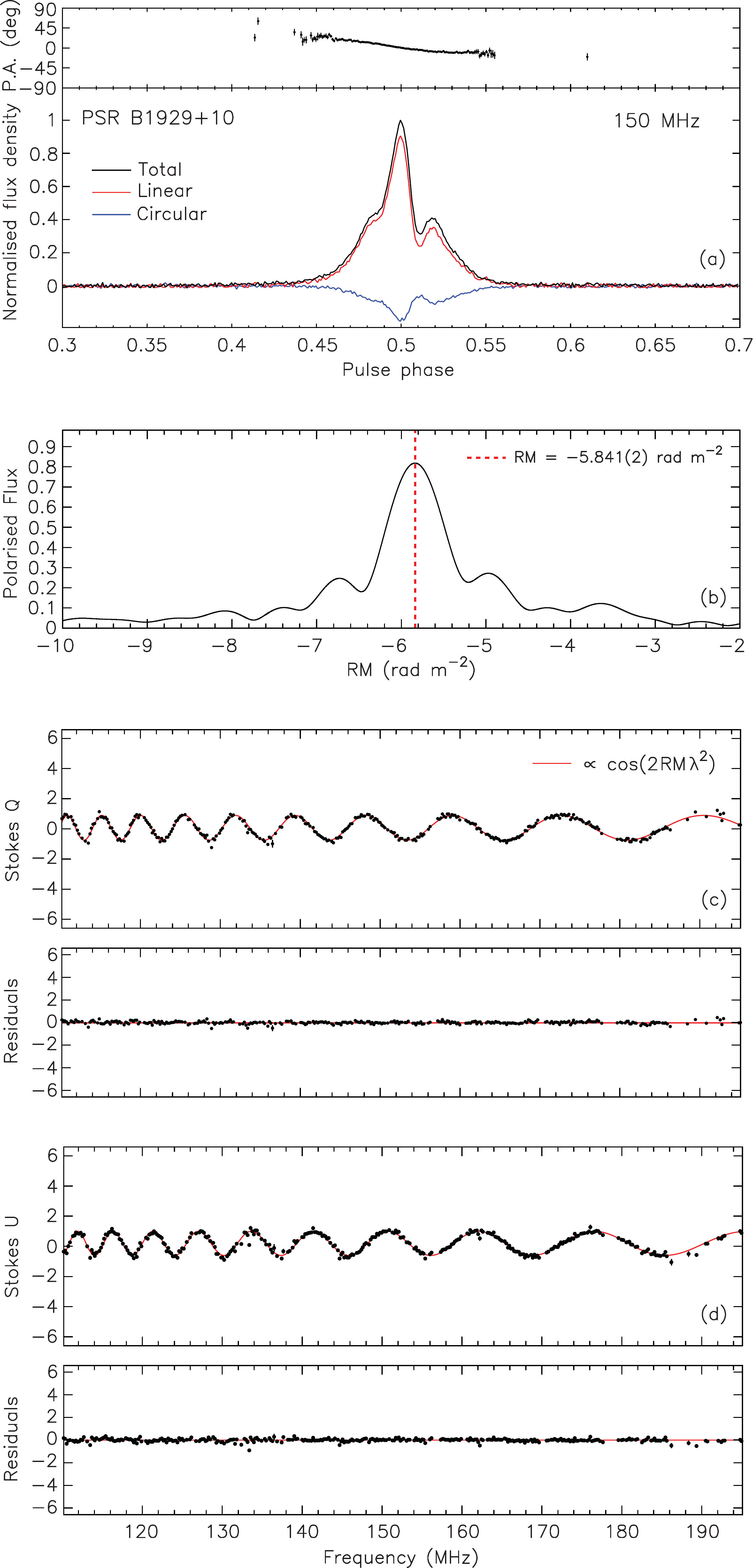}
\caption{(a) Average polarisation profile of PSR B1929+10 at 150 MHz. The black line shows the total intensity, the red line shows the linearly polarised intensity and the blue line, the circularly polarised intensity. The top panel shows the profile of the polarisation angle. (b) RM spectrum of PSR B1929+10, derived from the application of the RM Synthesis method to the polarisation data taken at 150 MHz. The maximum peak in the spectrum, indicated with a dashed red line, corresponds to the RM of the pulsar, ${\rm RM}=-5.837(2)$ rad m$^{-2}$, where the number in parentheses shows the uncertainty on the last significant digit. (c,d) Variation of the Stokes Q and U parameters across the observing band (black points), due to Faraday rotation, for the observation of PSR B1929+10. The red line shows the theoretical sinusoidal function of Q and U, assuming the determined RM value for this pulsar. The gaps in the frequency coverage are due to flagging of subbands that were affected by RFI. Below (c) and (d), the residuals from the subtraction of the theoretical function from the data are shown.}
\label{fig:B1929faraspec}%
\end{figure}

An important aspect of the quality of polarisation calibration is the stability of the pulse profiles as a function of hour angle. Before we can draw conclusions on the polarisation features of pulse profiles from LOFAR and how these compare to published profiles in the literature, we have to make sure that these are invariant with the time of observation. To that purpose, we have made a direct comparison between the PA, $L$ and $V$ profiles of PSR B2217+47 at transit (corresponding to $84^\circ$ elevation) and those at 5 hours prior and post transit (corresponding to $\approx 45^\circ$ elevation). Prior to comparing the profiles and calculating their residual differences, we phase-aligned the profiles either side of the transit to that at transit. This was done by determining the relative phase shifts between the profiles for which the $\chi^2$ between the total-intensity profiles is minimised. The polarisation profiles corresponding to the 3 pointings tested are shown on the same scale in Fig.~\ref{fig:B2217profdiffs}. Below the profiles of $L$ and $V$, we also show the residual difference as a function of pulse phase. It should be stressed that we have not attempted to perform any form of absolute polarisation calibration, as this would require a reference polarised signal of precisely known properties. Therefore, the PA profiles shown in Fig.~\ref{fig:B2217profdiffs}a and all other PA profiles from LOFAR, shown in this paper, are not meant to reflect the intrinsic angles of the polarised emission.

As was suggested earlier, the linear polarisation remains constant to within 6\%. The similarity between the PA profiles also suggests that the model largely corrects for the effects of parallactic rotation. The circular-polarisation profile, on the other hand, shows a much more significant variation between the three examined directions. For the chosen hour angles, the residuals are as large as 30\% of the circularly polarised flux at transit. This reaffirms the conclusions drawn from Fig.~\ref{fig:B2217beamtests}e.

Overall, our long-track observations of PSR B2217+47 show that beam calibration removes the strong dependence of the measured polarised flux on observing direction, at least for source elevations of $\gtrsim 30^\circ$. In that elevation range, we estimate that polarisation leakage is of the order of 5--10\%. However, we have observed an asymmetry in the performance of the beam model, mainly in the circularly polarised flux. The latter appears to increase by up to 75\% relative to its value at transit, in low-elevation ($<45^\circ$) observations towards the east. This is not well-understood, but it could be related to the topology of the horizon surrounding the Effelsberg station: the surrounding hills reach their maximum elevation of $\approx 30^\circ$ towards the south-east, which means that the signal is possibly contaminated by ground emission and possibly secondary reflections. 

In conclusion, our test observations have shown that the beam model is serviceable to within 5--10\% of systematic uncertainty, in observations with $>30^\circ$ source elevation. The polarisation data presented in this paper come from 15 observations with elevations of $>45^\circ$, whereas the remaining five observations were between 30 and 45$^\circ$. As such, we deem the polarisation properties of the pulsars presented here to be reliable to within the above systematic uncertainty. In the following, we have not tried to fold the systematic uncertainties arising from the beam model into the statistical uncertainties, but the reader should bear in mind that depending on the particularities of the observation, the quoted values could be different from the true values by the above percentages.

\begin{figure*}[!t]
\centering
\includegraphics[height=\dimexpr \textheight - 10\baselineskip\relax,width=\textwidth,keepaspectratio]{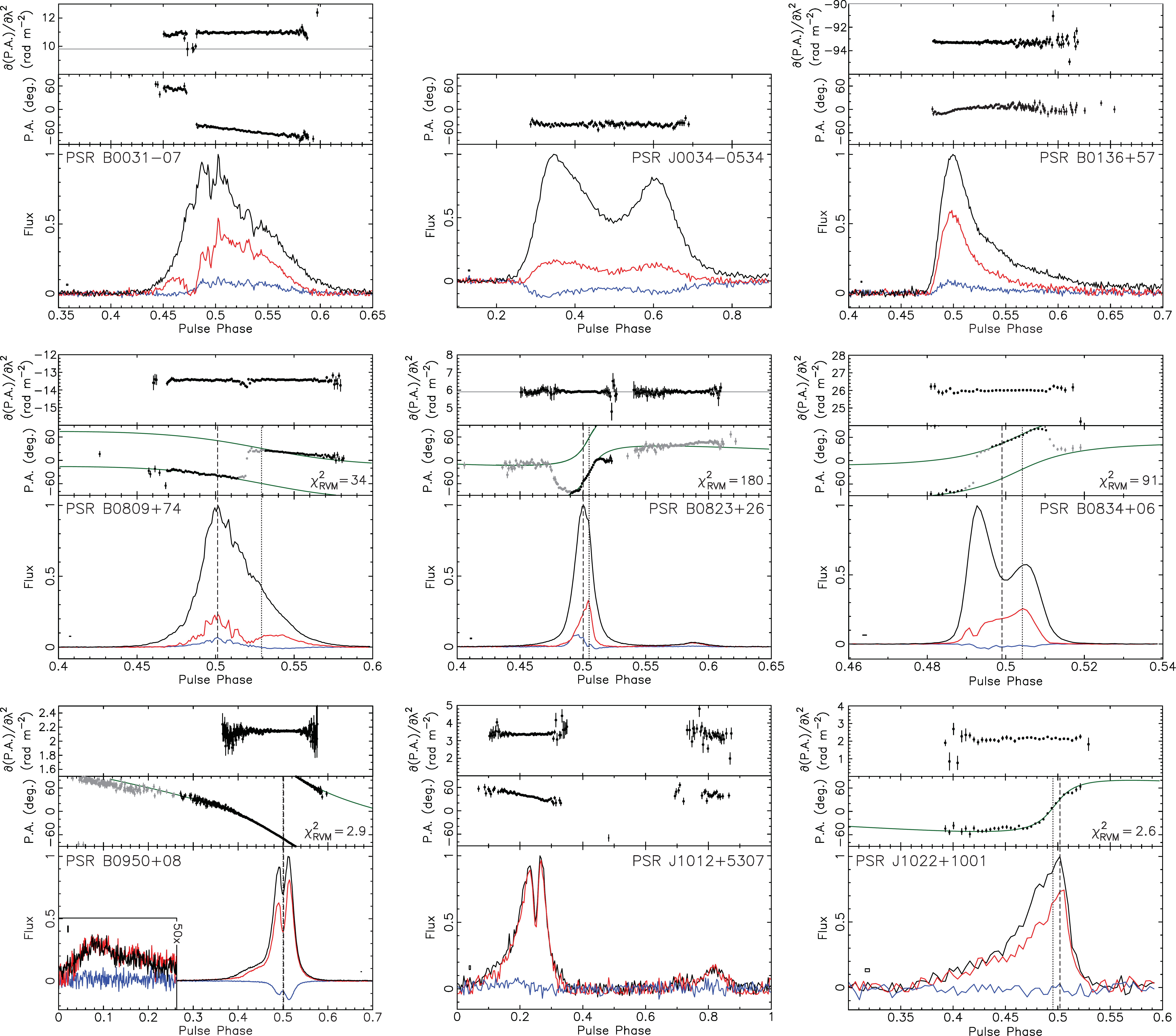}
\caption{The polarisation profiles of 20 pulsars observed with the LOFAR core at $\sim 150$ MHz. In each profile, the flux density of the total (black lines), linearly polarised (red lines) and circularly polarised emission (blue lines) is shown in arbitrary units normalised to unity. Above each flux profile, the profile of the polarisation angle (PA) of the linearly polarised emission is shown. Only PA values corresponding to phase bins having a signal-to-noise ratio in linear polarisation of $({\rm S/N})_L>3$ are shown. Above the PA profiles, we show the values of the first derivative of the PA with respect to $\lambda^2$, calculated within our band, at the phases of the PAs with $({\rm S/N})_L>5$ (see section~\ref{sec:rmvars}). As a reference, the published value of RM is indicated with a dashed, grey line, when it resides inside the plotted range. For PSRs J0034$-$0534 and B2111+46, we have not detected linear polarisation of astrophysical origin (see section~\ref{subsubsec:mspj0034}). For 15 pulsars, the phase of the emission assumed to be generated nearest to the magnetic pole is shown with a vertical, dashed line. The phase at the steepest PA gradient is shown with a vertical, dotted line, determined from RVM fits to the PAs (green lines). For some pulsars, the PA values shown in grey were zero-weighted to improve the RVM fit (see Section~\ref{sec:emheights} for details). The reduced $\chi^2_{\rm RVM}$ of each RVM fit is shown for each profile. The temporal resolution and the off-pulse RMS of each profile are shown near the bottom, left corner of the pulse profiles, with a square of corresponding dimensions. Finally, the weak interpulse of PSR B0950+08 is shown in the inset box, magnified $50\times$.}
\label{fig:polprofiles}%
\end{figure*}

\begin{figure*}[!t]
\ContinuedFloat
\centering
\includegraphics[height=\dimexpr \textheight - 10\baselineskip\relax,width=\textwidth,keepaspectratio]{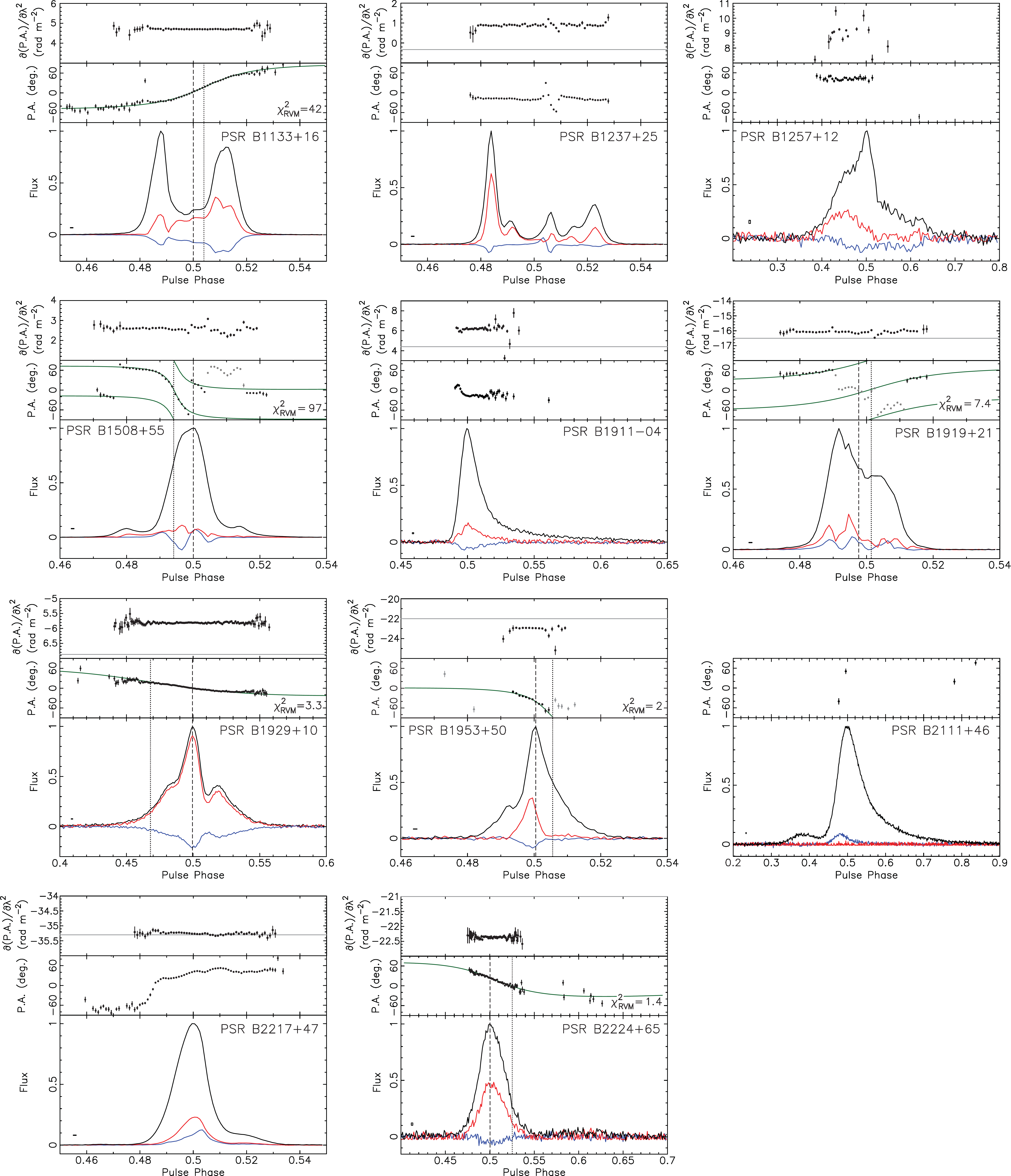}
\caption{Continued.}
\label{fig:polprofiles}%
\end{figure*}

\subsection{Faraday rotation}
A dominant effect that alters the polarisation properties of the pulsar signal, as detected on Earth, is Faraday rotation in the ISM. In addition, the Earth's ionosphere also causes Faraday rotation: typical ionospheric Faraday rotation contributes less than 1 rad m$^{-2}$ to the observed RM using the LOFAR stations, depending on the time of day, the season and Solar activity (Sotomayor-Beltran et al.~2013\nocite{ssh+13}). As a result of the geomagnetic field's polarity, ${\rm RM}_{\rm iono}$ is positive in the northern hemisphere and negative in the south. Faraday rotation causes the rotation of the linear polarisation, defined by the complex vector, $\tilde{P}=Q+U{\rm i}$, as the polarised electromagnetic waves propagate through the magnetised ISM. The amount of Faraday rotation to which the signal is subjected is proportional to the square of the wavelength of the emission, $\lambda^2$, and to the rotation measure, RM. For a given pulsar at a distance, $d$, the RM is proportional to $\langle n_eB_{\parallel}\rangle d$, where $n_{\rm e}$ is the free-electron density and $B_{\parallel}$ is the magnetic field projected along the line of sight, and the average is calculated over the distance to the pulsar. Depending on the average direction of the line-of-sight component of $B_{\parallel}$, the RM can be positive, when the field is directed towards the observer, or negative. The RM of a pulsar quantifies the magnitude of Faraday rotation towards the pulsar and it needs to be removed before calculating the amount of linear polarisation in average pulse profiles. Pulsar RMs can be accurately measured with the technique of RM Synthesis (Brentjens \& de Bruyn 2005)\nocite{bb05b}, whereby the sinusoidal variation of $Q$ and $U$ as a function of frequency is transformed into an RM spectrum (see Fig.~\ref{fig:B1929faraspec}). The maximum of the RM spectrum corresponds to the value of RM for which the magnitude of linear polarisation is maximised, and for which $\tilde{P}$ is completely de-rotated. According to Brentjens \& de Bruyn (2005), the statistical uncertainty on the determination of the maximum is calculated as $\sigma_{\rm RM}=0.5\times\Delta{\rm RM}/({\rm S/N})_L$, where $\Delta {\rm RM}$ is the full width at half maximum (FWHM) of the central lobe corresponding to the maximum of the RM spectrum, and $({\rm S/N})_L$ is the signal-to-noise ratio of the linearly polarised intensity in the average pulsar profile. The value of $\Delta {\rm RM}$ depends only on the total bandwidth in $\lambda^2$ of the observations, i.e.~$\Delta(\lambda^2)=\lambda_{\rm max}^2-\lambda_{\rm min}^2$: equivalently, $\Delta(\lambda^2)\propto (1/f_{\rm c}^2)\times B/(1-B^2)^2$, where $2B=\Delta f/f_{\rm c}$ is the fractional bandwidth. In our observations, the typical value was $\Delta {\rm RM}=2\sqrt{3}/\Delta (\lambda^2)\approx 0.6$ rad m$^{-2}$. 

The RM values that were used to correct for the Faraday rotation in our data are shown in Table~\ref{tab:polproperties}. For PSRs J0034$-$0534 and B2111+46 it was not possible to measure an RM due to the lack of measurable linear polarisation in our observations. Additionally, we have estimated the amount of Faraday rotation that was caused by the ionosphere, at the time of each observation, using the model of Sotomayor-Beltran et al.~(2013)\nocite{ssh+13}. The ionospheric RM, ${\rm RM}_{\rm iono}$, for each observation is shown in Col.~5 of Table~\ref{tab:polproperties}. We warn the reader that the measured Faraday rotation for each pulsar shown in Table~\ref{tab:polproperties} does not take into account the time- and direction-dependent Faraday rotation through the interplanetary and ionospheric magneto-ionic medium. It can be inferred from Table~\ref{tab:polproperties} that the contribution from the latter for our observations is of the order of 1 rad m$^{-2}$. The solar-wind contribution to the measured RMs is mainly dependent on the pulsar's angular separation from the Sun. You et al.~(2012)\nocite{ych+12} measured the solar-wind contribution to the RM of PSR J1022+1001, up to a separation of $\approx 20 R_{\odot}$, which they found was ${\rm RM}_{\odot}\sim 0.1$ rad m$^{-2}$. At the time of our observations, the separation between the pulsars and the Sun was $>45^\circ$, corresponding to $>200 R_{\odot}$, which implies that ${\rm RM}_\odot\ll 0.1$ rad m$^{-2}$. 

In conclusion, although the measurements can be considered accurate within the quoted statistical uncertainties --- as were calculated by the above analytic expression --- the quoted RM precision does not reflect our knowledge of the electron density and magnetisation of the ISM. In applications of pulsar RMs, e.g.~in studies of the Galactic magnetic field, systematic error estimation through models of the ionosphere and the solar-wind need to be also considered.

\section{Polarisation profiles at 150 MHz}
\label{sec:polprofs}
After the data-reduction and calibration process described in the previous section, we obtained time- and frequency-averaged polarisation profiles for 20 pulsars, at 150 MHz. All the calibrated profiles are shown in Fig.~\ref{fig:polprofiles}. Each plot shows a profile of the total flux density (black lines), normalised to unity. The linearly polarised and circularly polarised flux-density profiles are shown with red and blue lines, respectively. In the profiles shown, the pulse period of all non-recycled pulsars has been divided into 1,024 phase bins. In those cases, the temporal resolution of the profiles is in the range of $\sim 50$--$500$ $\upmu$s. For non-recycled pulsars this choice resolves the profile features adequately, while providing enough signal-to-noise for studies of the polarisation properties as a function of phase (see e.g.~Section~\ref{sec:rmvars}). For the MSPs PSR J0034$-$0534, J1012+5307, J1022+1001 and B1257+12, it was deemed adequate to use 256 phase bins across the profile, which corresponds to temporal resolution in the range of $\sim 5$--$50$ $\upmu$s. 

In the following sections, we will investigate the changes in the polarisation properties of the pulsars in our sample across a number of observing frequencies: namely, for most pulsars we supplemented the LOFAR data with archival observations at 400, 600, and 1400 MHz, taken with the Lovell radio telescope at Jodrell Bank (Gould \& Lyne 1998\nocite{gl98}; Stairs, Thorsett \& Camilo 1999\nocite{stc99}). For PSRs B0031$-$07, B0834+06 and B1919+21, observations with the GMRT, at 240 MHz, were also available (Johnston et al.~2008\nocite{jkm+08}). For a few of the pulsars presented in this paper, we did not use the archival profiles from Jodrell Bank, but instead employed higher-resolution polarisation profiles from other references. In particular this was the case for the 1400 MHz profile of PSR B1237+25, which came from observations with Arecibo (J.~Rankin, {\em priv.~communication}). In addition, the 1400 MHz profiles of the MSPs PSR J1012+5307 and PSR J1022+1001 came from observations by Xilouris et al.~(1998)\nocite{xkj+98}, with the Effelsberg telescope, and by Yan et al.~(2011)\nocite{ymv+11}, with the Parkes telescope, respectively.

For the purpose of presenting multi-frequency profiles in a way that allows direct comparison, we attempted to align the profiles across the different frequency bands mentioned above, based on components that were present and clearly identifiable at all frequencies. The alignment of profiles from different telescopes and observing systems is non-trivial because of significant changes in profile shape, unknown instrumental delays and uncertainties in interstellar dispersion (see e.g.~Hassall et al.~2012\nocite{hsh+12}). More specifically, for pulsars with simple profiles, containing a single dominant component at all frequencies (e.g.~PSR B2217+47), the alignment was based on that component. For pulsars with multiple components (e.g.~PSR B1237+25), the alignment was based on the mid-point of the profile. Finally, for some pulsars (e.g.~PSRs B2224+65 and J1022+1001) one or more components vanish above or below a certain frequency. For those, we identified the component that is common across all frequencies and aligned according to that. Finally, in order to extract as much information as possible from the LOFAR data, with regards to the frequency evolution of the polarisation profiles, we split the LOFAR band into 3 sub-bands, with centre frequencies of 120, 150 and 180 MHz.

\begin{figure}
\centering
\includegraphics[height=\dimexpr \textheight - 10\baselineskip\relax,width=0.27\textwidth,keepaspectratio]{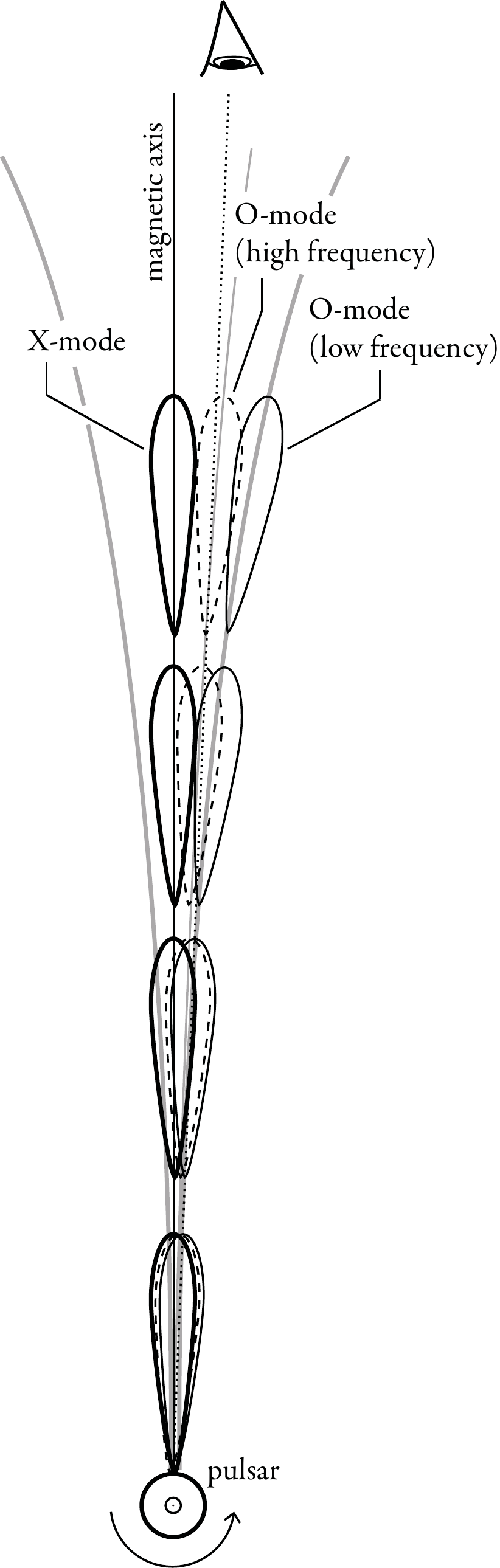}
\caption{Cartoon describing the effect of birefringence in radio-wave propagation through pulsar magnetospheres. In this representation, the reader is looking down the spin axis (circled dot). In the birefringent plasma, the X mode propagates parallel to the magnetic axis, unaffected by refraction, while the O mode follows the magnetic-field line direction (grey lines). Upon exiting the birefringent magnetospheric plasma, the two orthogonal propagation modes, the ordinary (O-mode) and extraordinary (X-mode), are beamed towards different directions. As the pulsar rotates, the two modes cross the observer's line of sight (dotted line) at different pulse phases. The degree of mode separation depends on the frequency of the emission: at high frequencies (dashed lobes), the separation between the modes is smaller than that at low frequencies (solid lobes).}
\label{fig:biref}%
\end{figure}

\section{Frequency evolution of pulsar polarisation}
\label{sec:freqevo}
\subsection{Introduction}
The spectra of pulsar polarisation across several octaves in frequency have been investigated in a number of studies (Morris, Graham \& Sieber~1981\nocite{mgs81}, hereafter MGS; Gould \& Lyne~1998\nocite{gl98}; Johnston et al.~2008\nocite{jkm+08}). In those studies, the results for several pulsars showed evidence for the occurrence of depolarisation with increasing observing frequency. In particular, MGS conclude that this effect is stronger for pulsars with long periods ($\gtrsim 300$ ms), whereas shorter period pulsars --- which at high frequencies are typically more highly polarised than pulsars with longer periods (von Hoensbroech, Lesch \& Kunzl~1998\nocite{hlk98})--- exhibit flatter polarisation spectra within the investigated frequency ranges ($\approx 200$ MHz--$8$ GHz). As yet, there is no consensus regarding the mechanism that is responsible for the observed polarisation behaviour as a function of frequency. Nevertheless, a number of models have been proposed that can explain the observed depolarisation at high frequencies (Ruderman \& Sutherland 1975\nocite{rs75}; Barnard 1986\nocite{bar86}; McKinnon 1997\nocite{mck97}; von Hoensbroech, Lesch \& Kunzl~1998\nocite{hlk98}). Many of these models are based on the birefringence of plasma in the open field-line region of pulsar magnetospheres. It is generally assumed that polarised radio emission is produced as the sum of two orthogonal propagation modes (OPMs), the ordinary (O) and extraordinary (X) mode (Manchester, Taylor \& Huguenin 1975\nocite{mth75}). 
The two modes are expected to be beamed in different directions, after exiting the birefringent medium, depending on the frequency-dependent refractive index: according to Barnard \& Arons (1986)\nocite{ba86}, the X mode propagates close to the magnetic axis, unaffected by the plasma, while the O mode is refracted along the magnetic-field lines. Hence, it is expected that the opening angle between the two modes increases towards low frequencies, where refraction is expected to be stronger (see Fig.~\ref{fig:biref}). At any given pulse phase, the observer's line of sight may cross both polarisation beams. The net orientation of the polarisation (O or X) will be that with the dominant intensity and the net intensity will be $L=|L_X-L_O|$. However, if both modes are beamed in roughly the same direction, as is the case for high-frequency emission, both modes have similar intensities which leads to weak or no net polarisation. The bifurcation of the emission could also be the explanation for discontinuities in the observed PA profiles of several pulsars. Changes of the PA by $\approx 90^\circ$, between adjacent phase bins, are commonly observed and sometimes referred to as orthogonal jumps (e.g.~Gangadhara 1997\nocite{gan97}). These may reflect the transition between the dominant orthogonal propagation modes. At the phases where the transition between the two modes occurs, if birefringence is the underlying process one should expect depolarisation due to the overlap of the beams.

If the above is true, it motivates the following observational tests that are a consequence of birefringence: (a) as the two propagation modes begin to overlap towards higher frequencies, the observed net polarisation of pulsars is expected to decrease with increasing observing frequency; and (b) the bifurcation of the beam due to birefringence implies that the width of the radio beam, and hence the observed pulse width, should increase with decreasing frequencies. An indication that the mechanism of birefringence operates in pulsar magnetospheres was provided by the work of McKinnon (1997), who studied the pulse broadening and depolarisation statistics for a few bright pulsars that exhibit simple profiles and timing behaviour. 

Besides birefringence, there are several other mechanisms that could operate in tandem. An additional complication arises because the range of altitudes over which the polarisation properties of radio waves are affected (e.g.~the path length over which refraction occurs) can also be frequency dependent (Barnard 1986); this is the so-called {\em polarisation limiting radius}. In addition, a number of studies assume that the different radio frequencies are generated at different heights ($r_{\rm em}$) above the pulsar surface, where $f\propto r_{\rm em}^{-3/2}$, the so called {\em radius-to-frequency mapping} (RFM; e.g.~Ruderman \& Sutherland 1975). In those studies, the main argument behind this assumption is the decreasing plasma density (and hence plasma frequency) as a function of altitude: i.e.~$\rho_{\rm e}\propto 1/r_{\rm em}^3$. We note, however, that the recent work by Hassall et al.~(2012), using simultaneous observations of pulsars from tens of MHz to $\approx 10$ GHz, concluded that at least for the pulsars studied the altitude of both low-frequency and high-frequency emission is confined within $\sim 100$ km. We will now focus on the observational predictions of birefringence.

\subsection{Polarisation fractions}
\label{subsec:polfracs}
According to McKinnon (1997)\nocite{mck97}, the impact of birefringence can be observationally traced by the increasing degree of linear polarisation with decreasing observing frequency {\em and}, at the same time, the inherent broadening of the integrated profile due to divergence of the orthogonal propagation modes towards low radio frequencies. McKinnon examined the frequency evolution of the linear-polarisation fraction using polarisation data between 150 MHz and 8 GHz, and the frequency evolution of pulse broadening between $\approx 20$ MHz and 10 GHz. In that work, the polarisation data at 150 MHz came mainly from the observations of Lyne, Smith \& Graham (1971)\nocite{lsg71} with the MkI 250-ft Jodrell Bank radio telescope. Owing to the limited sensitivity of that instrument, having a maximum bandwidth of 1 MHz, the polarisation fractions of only the brightest pulsars in that sample were measurable. In addition, the authors estimated the uncertainty on the polarisation fractions to be $\approx 10\%$, based on measurements of the relative sensitivity of the polarisation feeds.

We can re-investigate the above predictions of birefringence using polarisation data from LOFAR, complemented with polarisation profiles at higher frequencies. The LOFAR data correspond to subbands centred at 120, 150 and 180 MHz and the archival data, at 400, 600, and 1400 MHz, and where available 240 MHz. At each of those frequencies, we have calculated the fraction of linear polarisation using Eqs.~1--5. Some pulsars in our sample have multiple components, whose polarisation evolves significantly with frequency. An extreme example is PSR B2224+65, which has two clearly defined components above 400 MHz, separated by $\approx 0.1$ in phase. In the LOFAR band, the trailing component of this pulsar vanishes, whereas the persistent leading component exhibits its maximum polarisation fraction at those frequencies. The spectra of the polarisation fraction for each component of this pulsar are shown in Fig.~\ref{fig:polfractions}. In general, for such complex profiles a component-by-component analysis may be more appropriate but has not been attempted here.

In addition to the linearly polarised fraction, we have calculated the pulse width at each frequency as follows. Firstly, we calculated the cumulative flux-density distribution across the pulse period. Then, we calculated the pulse width as the phase interval containing a given fraction of the total pulse energy, by excluding a two-tailed percentage (left and right bound) from the cumulative flux-density distribution. This calculation was performed for a range of percentages between 0\% (corresponding to the entire pulse profile) and 100\% (corresponding to a pulse width equal to 0). The final value of the pulse width was the unweighted average of all the phase intervals. We have considered this approach as an alternative to the standard W10 or W50 --- corresponding to 10\% and 50\% of the profile's maximum flux density --- as radiometer noise and/or significant frequency evolution of pulse components in complex profiles often lead to erratic evolution of the pulse width as a function of observing frequency. Based on comparisons between our method and the more traditional approach, we concluded that over a range of hundreds of MHz of observing frequency, our method produced smoother evolution of the pulse width, even for complex, multi-component profiles, like that of PSR B1237+25. Fig.~\ref{fig:polfractions} shows the pulse width and the fraction of linear polarisation as a function of observing frequency, alongside the pulse profiles. 

We note that PSRs B0950+08 and B1929+10 have been known to possess a weak interpulse, separated from the main pulse by roughly half a period, detectable above 400 MHz (Gould \& Lyne 1998\nocite{gl98}). For PSR B0950+08, the weak interpulse is present in the LOFAR band (see Fig.~\ref{fig:polprofiles}). In Fig.~\ref{fig:polfractions}, we have included the profiles of the interpulse of this pulsar at the different frequencies. It can be seen that at all frequencies the interpulse is 100\% linearly polarised and is evidently much broader at 150 MHz, merging with the main pulse. For PSR B1929+10, we could not detect significant emission at the phase range where interpulse emission is seen above 400 MHz (see Fig.~\ref{fig:polfractions}).

Another interesting case is that of PSR B1237+25, which has been known to have two different modes of emission, the normal and abnormal (Lyne 1971\nocite{lyn71}). Hankins \& Rickett (1986)\nocite{hr86} observed this pulsar between 131 and 2380 MHz, and noted that during the observations, with integrations ranging between 10 and 60 minutes, the pulsar was in its normal mode. Since the typical time of our observations was $\approx 10$ minutes per source, we deemed unlikely that PSR B1237+25 switched between modes, especially given that it spends 85\% of the time emitting in the quiet-normal mode and a large fraction of the rest of the time, in the flare-normal mode; the quiet mode is quite rare (Srostlik \& Rankin~2005\nocite{sr05}). In addition, comparison of our profile with the one observed by Srostlik \& Rankin at 327 MHz (Fig.~3 in their paper) and that observed by Hankins \& Rickett at 131 MHz (Fig.~1c in their paper) shows that indeed the average total intensity and polarisation profile are very close to what we observe at 150 MHz. The higher-frequency profiles in Fig.~\ref{fig:polfractions} of this pulsar follow the evolution seen by Hankins \& Rickett, where the ratio of the first leading component over the last trailing component decreases with frequency. In addition, the abnormal mode of this pulsar is associated with flaring of the core component, which we do not see in any of the profiles. Therefore, although it is not explicitly mentioned in Gould \& Lyne (1998), we favour that the profiles of this pulsar at 400 and 600 MHz in Fig.~\ref{fig:polfractions} show normal-mode emission. The normal-mode profile at 1400 MHz, in the same figure, was taken from observations by J.~Rankin with Arecibo (priv. communication).

\begin{figure*}
\centering
\includegraphics[height=\dimexpr \textheight - 10\baselineskip\relax,width=\textwidth,keepaspectratio]{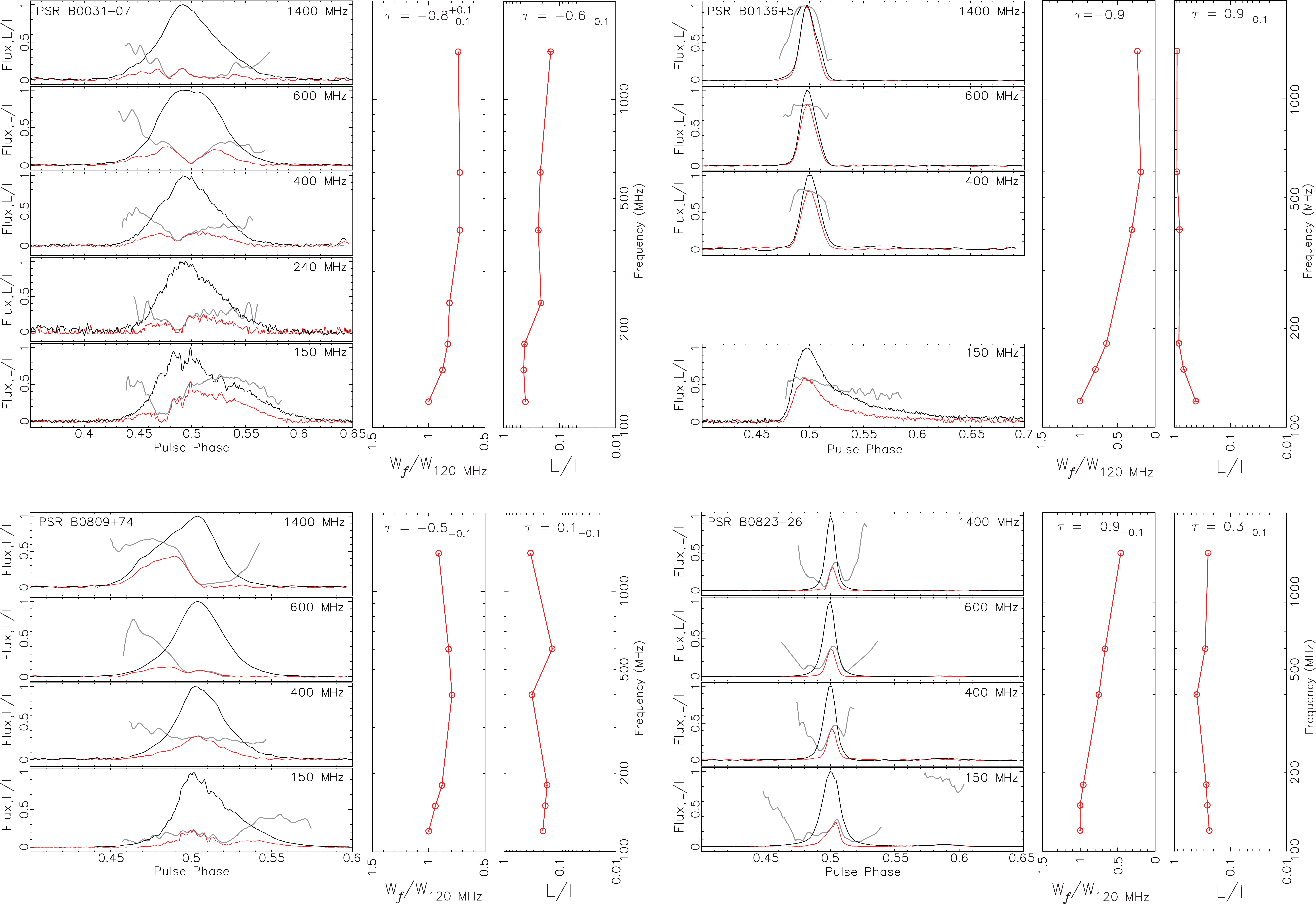}
\caption{Profiles of the total (black lines) and linearly polarised intensity (red lines) at 150 MHz, from this work, and at 400, 600, and 1400 MHz, from archival observations with the Lovell telescope, for each of the 16 non-recycled pulsars studied in this paper. The fraction of linear polarisation ($L/I$) as a function of phase is shown with grey lines. The $L/I$ profiles have been smoothed in phase using the method described in \ref{subsec:DataAnalysis}. All profiles have been roughly aligned in phase. Where available, the profile at 240 MHz from archival observations with the GMRT is also shown. The profile of PSR B1237+25 at 1400 MHz came from observations with Arecibo (J.~Rankin, {\em priv. communication}). The pulse width, normalised by its value at 120 MHz ($W_f/W_{120 \ {\rm MHz}}$), and the fraction of linear polarisation ($L/I$) are plotted as a function of observing frequency alongside the profiles. For the purpose of increasing the information on the frequency evolution of the pulse width and the  polarisation fraction, the LOFAR band has been split into three 30 MHz subbands. In those plots we show the value of Kendall's $\tau$, which is a measure of the correlation (positive value) or anti-correlation (negative value) of the plotted quantity with observing frequency. The asymmetric uncertainties on $\tau$ have been derived from a large number of Monte Carlo realisations of the data, assuming Gaussian statistics; only uncertainties of $\geq 0.1$ are shown. For PSR B0950+08, the 3 distinct components seen in the linear polarisation profile are marked with letters L, C and T, corresponding to the leading, central and trailing component, respectively. At 1400 MHz, component L is very weak at the phase where it is clearly visible at 600 MHz (marked with grey for reference). Above each flux-density profile of PSR B0950+08, also shown is the corresponding PA profile to aid the discussion in section \ref{subsec:b0950}.}
\label{fig:polfractions}%
\end{figure*}

\begin{figure*}
\ContinuedFloat
\centering
\includegraphics[height=\dimexpr \textheight - 10\baselineskip\relax,width=\textwidth,keepaspectratio]{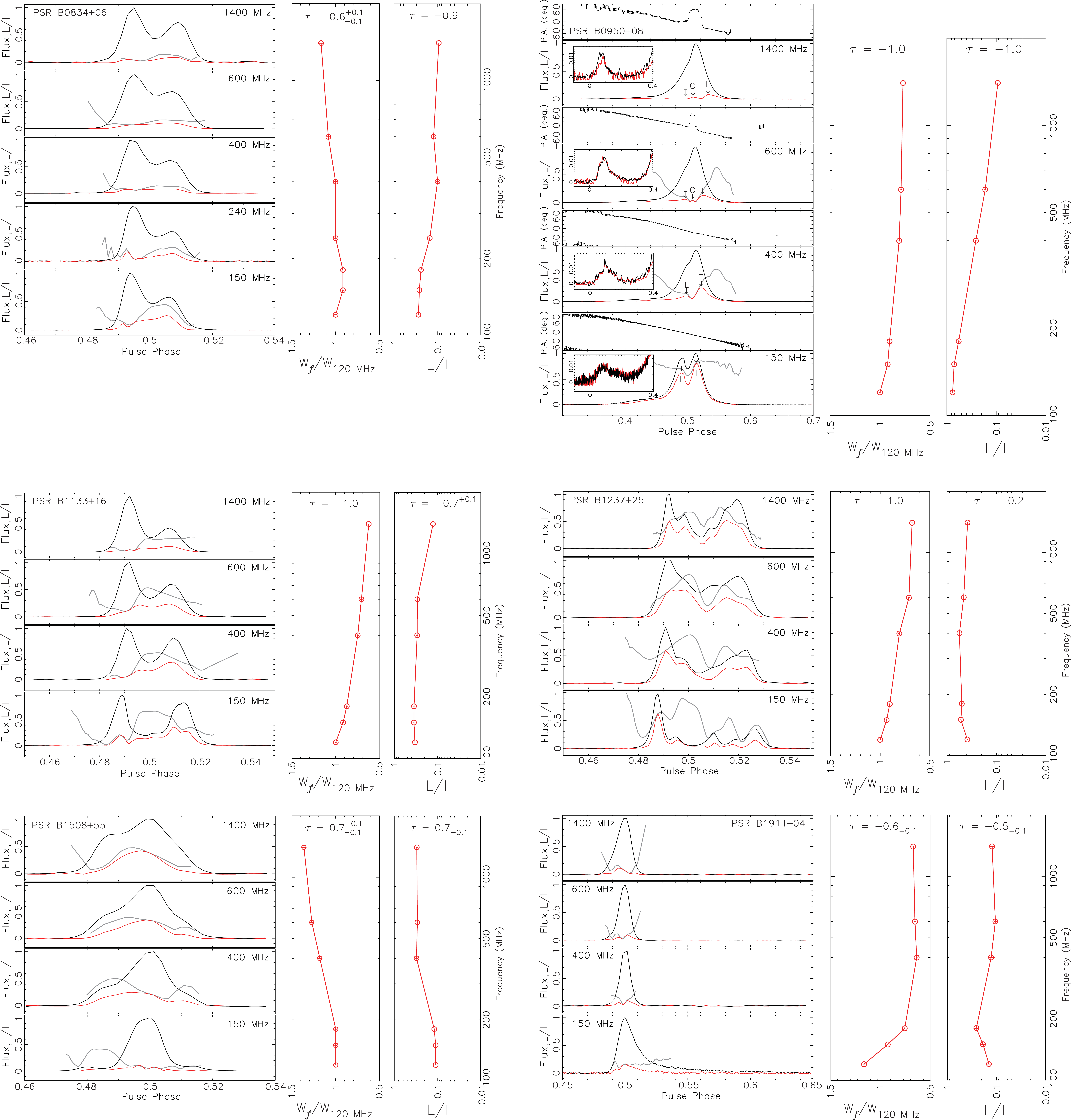}
\caption{Continued.}
\label{fig:polfractions}%
\end{figure*}

\begin{figure*}
\ContinuedFloat
\centering
\includegraphics[height=\dimexpr \textheight - 10\baselineskip\relax,width=\textwidth,keepaspectratio]{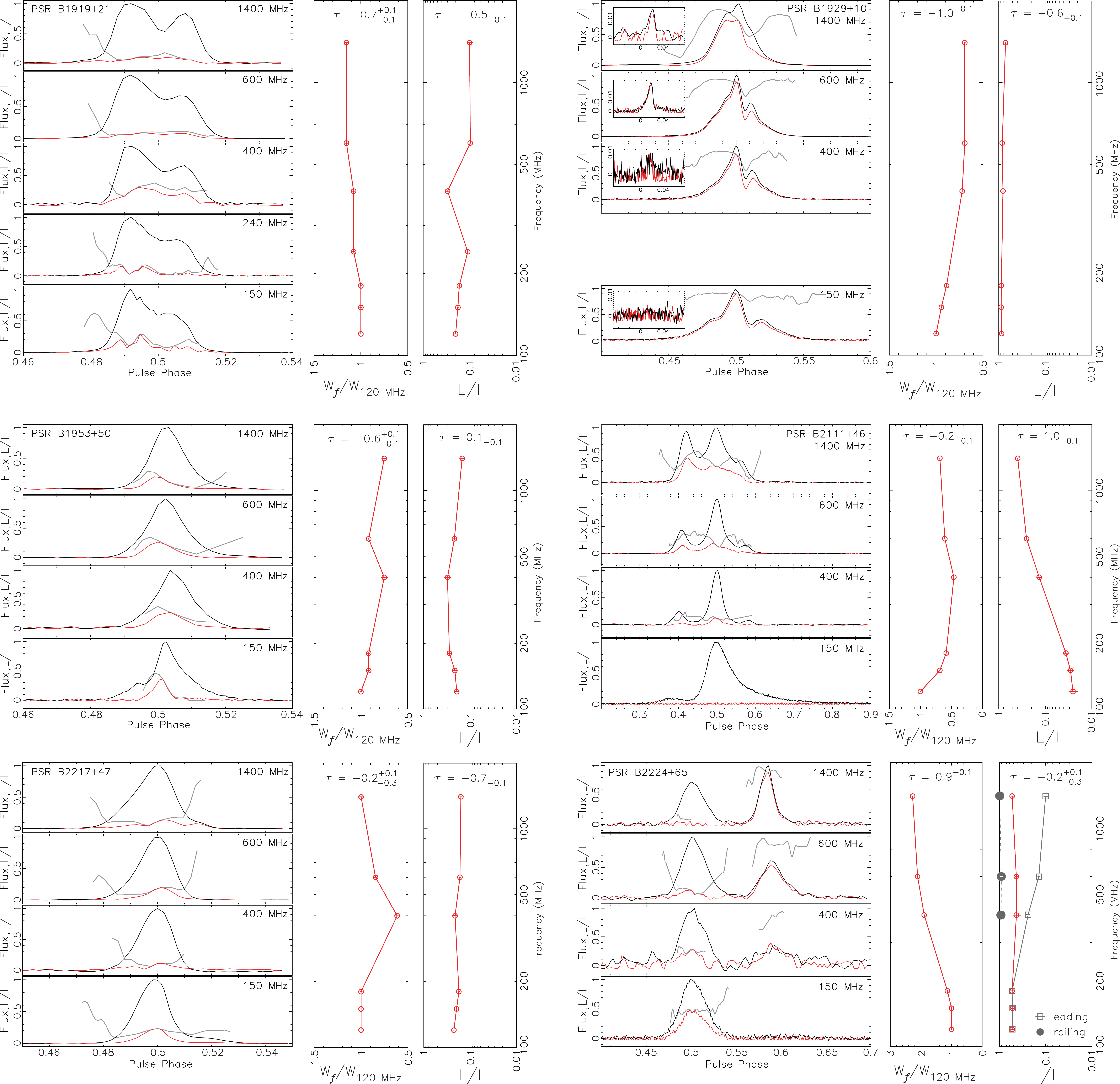}
\caption{Continued.}
\label{fig:polfractions}%
\end{figure*}

A reliable measure of the type and degree of correlation between two quantities that are ordered across a range is Kendall's $\tau$. This test is non-parametric and does not assume, e.g.~that there is a linear relation between the quantities. In each of the plots of Fig.~\ref{fig:polfractions}, we show the value of Kendall's $\tau$, calculated between the observing frequencies and the pulse width or the linear polarisation fraction. In order to take into account the uncertainties on the above quantities at each frequency, we calculated $\tau$ for a large number of realisations of the pulse width and linear polarisation fraction, assuming their uncertainties are Gaussian. The $1\sigma$ asymmetric uncertainties drawn from the Monte Carlo distribution of $\tau$ are also shown next to the value of $\tau$: where the uncertainty is $<0.1$, it is not shown. Kendall's $\tau$ takes values between $-1$ and $1$, with the extreme negative values implying negative correlation and extreme positive values, positive correlation. So, according to McKinnon (1997), we should expect to see a negative correlation between pulse width and frequency and polarisation fraction and frequency, due to birefringence. However, out of the 16 pulsars of our sample, only 9 show clear pulse broadening towards low frequencies (i.e.~have $\tau<-0.5$). Similarly, only 6 pulsars exhibit a decrease in the fraction of linear polarisation towards high frequencies. In contrast, we see that four pulsars show pulse broadening towards high frequencies ($\tau>0.5$) and three pulsars show increasing fractional polarisation with frequency. Upon closer inspection of the pulse profiles of Fig.~\ref{fig:polfractions}, we see that PSRs B0136+57 and B2111+46 are significantly scattered by the ISM at LOFAR frequencies. This could explain the positive correlation between frequency and polarisation fraction, as well as the pulse broadening at low frequencies. If we exclude those two pulsars from our investigation, we still see that approximately 60\% of the pulse-width distributions are consistent with birefringence, while only about 30\% show a positive correlation between pulse width and frequency. Finally, after excluding PSRs B0136+57 and B2111+46 on the basis that their profile evolution with frequency below 400 MHz is clearly dominated by scattering, it is interesting to note that only PSR B1508+55 of all pulsars examined becomes intrinsically more polarised towards high frequencies.

In conclusion, the frequency evolution of neither the pulse width nor the degree of linear polarisation show conclusively the effects of birefringence, in the small sample of pulsars examined. The distributions of $\tau$ for the pulse widths show some indication of pulse broadening towards lower frequencies, even after excluding the clearly scattered pulsars, but we also see cases where the opposite is true. On the other hand, the distribution of $\tau$ for the linear polarisation fractions is even more balanced between cases that support birefringence and those that do not. We have performed a KS test between each of two observed distributions of $\tau$ and a theoretical distribution where all values of $\tau$ are uniformly distributed in $\tau\leq -0.5$ (Massey 1951)\nocite{mas51}. We find that the probability that the observed and theoretical pulse width distributions are related is $0.04^{+0.1}_{-0.03}\%$. Similarly, we find that the probability that the observed distribution of linear-polarisation fractions is related to the theoretical is only $0.003^{+0.01}_{-0.002}\%$. These values reaffirm our conclusion that there is no strong evidence for birefringence in the data.

\section{Effects of scattering on the frequency evolution of the PA}
\label{sec:rmvars}
\subsection{Introduction}
The effects of scattering on pulsar polarisation have been discussed in several publications (Komesaroff, Hamilton \& Ables 1972\nocite{kha72}; Li \& Han 2003\nocite{lh03}; Noutsos et al. 2009\nocite{nkk+09}; Karastergiou 2009\nocite{kar09}). The main conclusion from these studies was that scattering can cause flattening of the PA profiles, and in some cases smear away orthogonal jumps. More recently, a secondary effect attributed to scattering was detected in polarisation data obtained at 1400 MHz with the Parkes telescope (Noutsos et al. 2009\nocite{nkk+09}). In those data, it was seen that for a number of high-DM pulsars (${\rm DM}\gg 100$ pc cm$^{-3}$) the amount of PA rotation across the band varies significantly as a function of pulse phase. For example, the highest peak-to-peak variation of $100$ rad m$^{-2}$ was observed for PSR J1056$-$6258, which was the pulsar with the second highest DM in that sample. Furthermore, it was shown that within the errors the PA rotation is consistent with Faraday rotation, irrespectively of the choice of pulse phase. If interpreted as Faraday rotation, it would seem that the RM of those pulsars varies as a function of pulse longitude. However, in those studies it was suggested, based on the positive correlation of the magnitude of those variations with pulsar DM, and was also independently shown through simulations, that this is an artefact of scattering and is physically independent of Faraday rotation (Noutsos et al.~2009\nocite{nkk+09}; Karastergiou 2009\nocite{kar09}). 

In the following sections, we present arguments that strengthen the case for scattering being responsible for the so-called phase-resolved RM variations. As such, we have refrained from referring to the phase-resolved rotation of the PA as a function of $\lambda^2$ as RM. Instead, we have used the following notation: if $\Psi(\phi,\lambda^2)$ (hereafter just $\Psi$) is the PA as a function of pulse phase, $\phi$, and $\lambda^2$, then $\Psi_{\lambda^2}$ and $\Psi_{\phi}$ are the first-order partial derivatives of $\Psi$ with respect to $\lambda^2$ and $\phi$, respectively. In addition, $\Psi_{{\lambda^2}\phi}$ is the second-order mixed partial derivative of $\Psi$ with respect to $\lambda^2$ and $\phi$.

\subsection{Toy model}
If scattering is indeed the reason for the apparent variations of $\Psi_{\lambda^2}$ as a function of $\phi$, then its effect can be explained as follows. At any given pulse phase, $\phi_0$, the polarised intensity of the scattered profile, $\tilde{P}(\phi_0)$, is the result of the convolution of the intrinsic polarised intensity, $\tilde{p}(\phi)$, with an unknown scattering function. Under the assumption that scattering is caused by a thin screen located at a distance that is much smaller than that of the pulsar, the scattering function can be approximated with a one-sided exponential of characteristic timescale, $\tau_{\rm s}$ (Cronyn 1970\nocite{cro70}; Lee \& Jokipii 1975\nocite{lj75}). Hence, the change in polarised intensity at phase $\phi_0$, due to thin-screen scattering, is given by 
\begin{equation}
\label{eq:scat}
\tilde{P}(\phi_0)=\frac{1}{\tau_{\rm s}}\int_{0}^{\phi_0}\tilde{p}(\phi){\rm e}^{-(\phi_0-\phi)/\tau_{\rm s}}{\rm d}\phi.
\end{equation}
The normalisation factor, $1/\tau_{\rm s}$, ensures that pulse energy is conserved between the intrinsic and scattered profiles. In the simple case where the scattered radiation has a Gaussian angular intensity distribution, $\tau_{\rm s}\propto \lambda^4$ (Cronyn 1970\nocite{cro70}). Therefore, it can be seen that the range of phases over which scattering has a measurable effect is strongly dependent on frequency. As a result of the aforementioned convolution, large changes of the PA in the intrinsic polarisation profile --- such as steep PA gradients and/or orthogonal jumps --- are observed as smaller changes, in the scattered profile. In such a scenario, the reported variations of $\Psi_{\lambda^2}$ as a function of $\phi$ are a direct consequence of the frequency dependence of $\Psi_{\phi}$, due to scattering. 

In Appendix A, we show that the effects of scattering on steep PA profiles and the frequency evolution of the PA can be estimated using a simple polarisation profile that has been scattered (but not Faraday-rotated) by a thin screen. The main results from our simple model are as follows. (a) As has been discussed in previous work, scattering reduces the steepness PA profiles, with the effect being greater at lower frequencies; we find that this effect is also variable with pulse phase. (b) The value of $\Psi_{\lambda^2}$ (normally a measure of Faraday rotation) is not constant with pulse phase but varies across the profile. In other words, we find that, for a given pulse phase, scattering can indeed introduce a change of the PA with frequency that is indistinguishable from Faraday rotation. (c) Finally, the most interesting result is that the maximum values of $\Psi_{{\lambda^2}\phi}$ in the profile are exactly proportional to $1/\lambda^2$. In other words, we find that gradients of $\Psi_{\lambda^2}$ are expected to be steeper at higher frequencies.

\subsection{Data analysis}
\label{subsec:DataAnalysis}
\begin{figure}
\centering
\includegraphics[height=\dimexpr \textheight - 10\baselineskip\relax,width=0.47\textwidth,keepaspectratio]{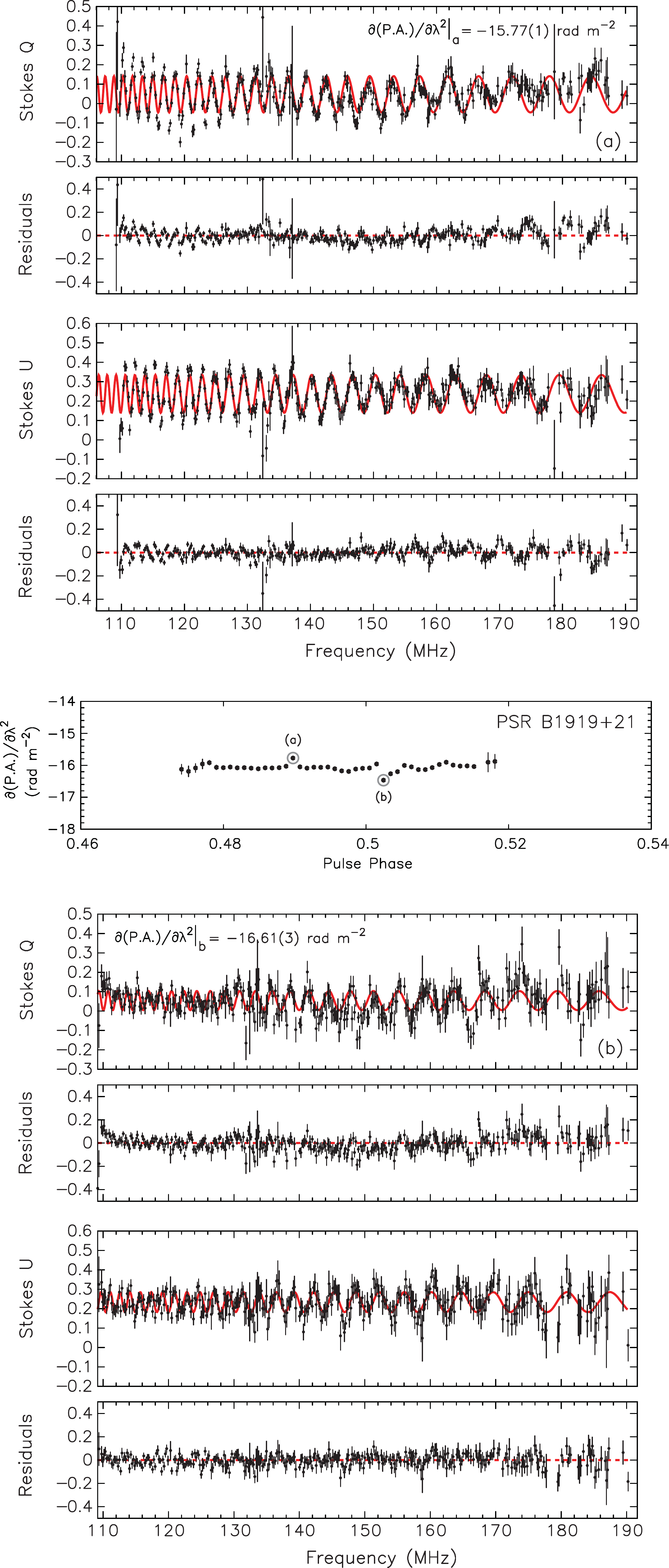}
\caption{Scatter plots of the variation of the Stokes Q and U parameters across the HBA band, calculated for the two pulse phases of PSR B1919+21 that correspond to (a) the maximum ($-15.77(1)$ rad m$^{-2}$) and (b) the minimum value ($-16.61(3)$ rad m$^{-2}$) of the first derivative of the PA with respect to $\lambda^2$ (middle plot; the selected phases are highlighted with grey circles). The expected periodicity of the Stokes parameters with frequency, based on the above values of $\partial{\rm PA}/\partial\lambda^2$, is shown for each case with a red curve. The residual difference between the expected variation and the data is plotted below each of the Stokes Q and U plots, where the dashed red line corresponds to zero residual difference.}
\label{fig:B1919phaserms}%
\end{figure}

The last conclusion from our simple toy model is perhaps unexpected: it implies that, if scattering is responsible for the variations of $\Psi_{\lambda^2}$ as a function of phase, as have been observed at 1400 MHz, then the typical magnitude of these variations should be $\sim 100$ times lower at 150 MHz. This prediction motivates us to investigate this effect at LOFAR frequencies and compare it with the published data at 1400 MHz. In order to detect the presence and quantify the magnitude of variations of $\Psi_{\lambda^2}$ across the LOFAR profiles, firstly we performed the technique of RM Synthesis on the Stokes Q and U signals of every phase bin across the polarisation profiles of Fig.~\ref{fig:polprofiles}. The resulting profiles of $\Psi_{\lambda^2}$ are shown in the same figure, above each PA profile. Furthermore, we checked how well the data in the outliers of those profiles follow the expected dependence on $\lambda^2$ (the main assumption of RM Synthesis) by examining the corresponding Stokes Q and U values as a function of frequency. In Fig.~\ref{fig:B1919phaserms}, we show the variation of the Stokes parameters across the HBA band, for two phase bins of the profile of PSR B1919+21, corresponding to the minimum and maximum significant value of $\Psi_{\lambda^2}$. Despite the low S/N per channel and the baseline and amplitude variations across the band, which are evident in the residual difference from the expected function shown with the red curves, it is clear that the data track well the expected periodicity as a function of frequency. Hence we can be confident that, even for those phase bins, the $\lambda^2$ dependence is the correct assumption.

Secondly, we selected only those pulsars that show hints of variations of $\Psi_{\lambda^2}$, across the pulse, based on visual inspection: these were PSRs B0031$-$07, B0809+74, B0823+26, B0834+06, B1237+25, B1508+55, B1919+21 and B2217+47. Subsequently, for the selected pulsars we elected to measure $\Psi_{{\lambda^2}\phi}$ as a function of pulse phase across the respective profiles. However, differentiation of noisy, unevenly sampled data, such as the profiles of $\Psi_{\lambda^2}$, is a well-known problem (Cullum~1971\nocite{cul71}; Ruzmaikin et al.~1988\nocite{rss88}). A general solution typically followed in the literature is to describe the data with a smooth function (e.g.~a polynomial or a cubic spline). More specifically, for the purposes of differentiating noisy digital signals, Savitzky and Golay popularised a smoothing algorithm that uses least-squares fitting of a low-degree polynomial to subsets of the data set in question (Savitzky \& Golay 1964\nocite{sg64}). The polynomial equations describing the subsets can be solved simultaneously to provide a single set of convolution coefficients that can be multiplied with the noisy signal to yield a smooth function across the entire data set. A requirement of the Savitzky--Golay (SG) differentiation filter is that the data are uniformly sampled across the application range. Our $\Psi_{\lambda^2}$ profiles are often uneven due to the imposed limits on signal-to-noise, which means that only phase bins with ${\rm S/N}>5$ were considered. Nevertheless, it is possible to perform a linear interpolation that will ensure uniformity. Following the interpolation, we applied a 4th-order SG filter to the $\Psi_{\lambda^2}$ profiles, operating on five neighbouring data points either side of each data point of the profile. To avoid boundary problems, the first and last five data points (of the interpolated profile) were ignored in the final calculations. 

The above procedure yielded a smooth function of $\Psi_{\lambda^2}$ and its first-order derivative with pulse phase, for each pulsar. From that, we obtained the maximum value of $|\Psi_{{\lambda^2}\phi}|$, for the eight selected pulsars at 150 MHz. The described procedure was repeated for a different sample of nine pulsars observed at 1400 MHz by Noutsos et al.~(2009)\nocite{nkk+09}, for which significant variations of $\Psi_{\lambda^2}$ with pulse phase were measured. In total, our analysis produced eight values of ${\rm max}(|\Psi_{{\lambda^2}\phi}|)$ at 150 MHz, with a weighted median of $12_{-5}^{+32}$ m$^{-2}$, and nine values at 1400 MHz, with a weighted median of $816_{-582}^{+291}$ m$^{-2}$. Assuming that the uncertainties on each value of ${\rm max}(|\Psi_{{\lambda^2}\phi}|)$ are Gaussian, it then follows that between 150 and 1400 MHz ${\rm max}(|\Psi_{{\lambda^2}\phi}|)\propto \lambda^{-1.7(5)}$. This result is consistent within the $1\sigma$ uncertainty with the prediction of the toy model, albeit there is a large uncertainty, mainly due to the limited sample. In the future, it will be possible to increase the sample of pulsars for which this study can be made. Moreover, it will be possible to include data at 1400 MHz of the pulsars we have observed with LOFAR, so as to minimise the systematic uncertainties introduced by the different morphology of the profiles between different pulsars. Nevertheless, as we showed in Sections~\ref{sec:freqevo}, even when restricting this analysis to multi-frequency data of the same pulsars, one may still need to account for intrinsic profile evolution.

\section{Emission heights}
\label{sec:emheights}   
\subsection{Introduction}
In the framework of the Rotating Vector Model (RVM; Radhakrishnan \& Cooke 1969), the emission is assumed to originate from the pulsar surface. In this model, the oblique rotation of the pulsar's beam relative to the observer's line of sight is reflected by the parallactic rotation of the PA across the profile. The PA profile described by RVM resembles an S-curve (hereafter RVM swing), where the inflexion point corresponds to the phase at the closest approach of the observer to the magnetic pole, $\phi_0$. The exact shape of the PA depends on the angle between the spin and magnetic axes, $\alpha$, and the angle between the magnetic axis and the observer's line of sight at the closest approach, $\beta$. Equivalently, we can define the viewing angle of a distant observer with respect to the pulsar's spin axis, $\zeta=\alpha+\beta$. The general form of the RVM function gives the PA as a function of phase,
\begin{equation} 
\label{eq:rvmeq}
\Psi(\phi)=\Psi_0+\tan^{-1}\left[\frac{\sin\alpha\sin(\phi-\phi_0)}{\sin\zeta\cos\alpha-\cos\zeta\sin\alpha\cos(\phi-\phi_0)}\right],
\end{equation}
where $\Psi_0$ is the PA at $\phi_0$.

At $\phi_0$, the observer's meridional plane contains both the spin axis and the magnetic axis, and the rate of change of the PA with phase takes its maximum value, i.e.~$({\rm d}\Psi/{\rm d}\phi)_{\rm max}=\sin\alpha/\sin\beta$. However, radio emission is thought to be generated at a finite altitude above the polar caps, $r_{\rm em}>0$, by relativistic plasma accelerated along the dipolar field lines. In such a scenario, the pulsar's co-rotating magnetosphere --- as seen by the inertial observer --- contributes to the bending of the beam of accelerated particles. As a result, it introduces a lag between the phase of the location of the emission and the phase of the corresponding emission, $\Delta\phi$. In other words, the magnetic-field lines are bent forward in the direction of the pulsar rotation, so that emission that is generated by magnetic field at phase $\phi_0$, in the co-rotating frame, is observed earlier, at $\phi_{\rm em}$ (Blaskiewicz, Cordes \& Wasserman 1991, hereafter BCW\nocite{bcw91}). In addition, due to this effect the phase at the steepest PA gradient (PA inflexion), $\phi_0$, does not correspond to the closest approach of the observer's line of sight to the magnetic pole but is shifted to later phases. BCW showed that the total lag between $\phi_{\rm em}$ and $\phi_0$ can be approximated with
\begin{equation}
\label{eq:bcwlag}
\Delta\phi_{\rm BCW}=4\frac{r_{\rm em}}{R_{\rm LC}},
\end{equation}
where $R_{\rm LC}=cP/(2\pi)$ is the light-cylinder radius. One can use this expression to calculate $r_{\rm em}$, given that $\phi_0$ and $\phi_{\rm em}$ can be determined, as will be described in the next section. It is important to mention that, as Dyks (2008)\nocite{dyk08} noted, this is not an effect caused by beam aberration but simply by the co-rotation of the emission region as seen in the observer's reference frame.

\subsection{Data analysis}
\subsubsection{Determining $\phi_0$}
The determination of the phase corresponding to the PA inflexion ($\phi_0$) is typically based on the observed polarisation. To this aim, it is common to employ RVM fits to the data and identify $\phi_0$ as the phase at the steepest gradient in the PA profile. However, this procedure may be hampered by processes that are intrinsic or extrinsic to the pulsar. Consequently, such RVM fits can result in large uncertainties on $\phi_0$. One of the reasons is that for several pulsars we have incomplete polarisation information across the profile to obtain a reliable fit, perhaps because our line of sight samples only a small cross section of the pulsar's active regions. Independently of viewing geometry, it is also possible that the intrinsic polarisation of the pulsar is unevenly distributed across the open field-line region, so that given our instrument's sensitivity, a complete PA swing could be unobtainable. Moreover, polarised emission generated at different altitudes across the pulse can introduce distorting features to PA profiles (Hibschman \& Arons 2001\nocite{ha01}). Ramachandran \& Kramer (2003) suggested that the evident notch in the PA profile of PSR J1022+1001 at 1400 MHz could be due to such altitude-dependent polarisation (see Section~\ref{subsubsec:mspj1022}). The authors were able to fit two separate RVM curves to the PAs in the phase ranges either side of the notch. Finally, a number of propagation effects in the pulsar magnetosphere have been proposed that act towards modifying the shape of the observed PA profiles, e.g.~wave-mode coupling and the quasi-tangential propagation effect (Wang et al.~2010\nocite{wlh10}). In addition, as was explained in Section~\ref{sec:freqevo}, irregular features, like OPM jumps, could be the result of birefringence.

Besides the aforementioned intrinsic effects, a number of extrinsic processes can also distort the polarisation signal, as it is observed on Earth. For example, the deviation of pulsar PA profiles from an RVM swing can partly be due to the data-averaging process. Gil \& Lyne (1995)\nocite{gl95} and later Mitra et al.~(2007)\nocite{mrg07} showed that the individual pulses from PSR B0329+54 have polarisation that is consistent with an RVM swing. However, the average polarisation of this pulsar yields a PA profile that significantly deviates from that shape. In that work, it was seen that each of the single pulses is rather well confined to one of the two OPMs, so that the averaging process results in the PA profile that is determined by the relative strength and number of the single pulses at each pulse phase. Furthermore, as was discussed in Section~\ref{sec:rmvars}, the intrinsic PA shapes can be distorted by scattering through the ISM. This is independent of time-averaging, since changes in the ISM occur at much longer scales compared to the length of our observations, but it can strongly depend on frequency-averaging, due to the strong frequency dependence of scattering.
 
Despite the aforementioned shortcomings, we were able to determine the value of $\phi_0$ via fits of Eq.~\ref{eq:rvmeq} to the PA profiles of 11 pulsars. A small number of profiles at 150 MHz, like those of PSRs B1133+16 and B1929+10, have a simple PA evolution with phase that can be well described with an RVM curve. For others, like PSR B1237+25 and the MSPs B1257+12 and J1012+5307, the PA profile at 150 MHz is too flat and a reliable fit could not be obtained. In some cases, like those of PSRs B0809+74, B0834+06 and B1508+55, we needed to account for a number of OPM jumps across the profile, in order to obtain a good fit. More specifically, we allowed the PAs to rotate independently by 90$^\circ$, in order to minimise the $\chi^2$ of the fits. We stress that these jumps are a property of the emission and are independent of the viewing geometry. For a discussion of the used method to determine $\phi_0$ and its uncertainty we refer to Rookyard, Weltevrede \& Johnston (MNRAS, submitted). It is worth noting that the PA profile of PSR B2217+47 exhibits a swing at $\phi\approx 0.485$, across which the PA changes by $\approx 90^\circ$; this could be mistaken for an RVM swing. However, a closer look at the behaviour of the linear polarisation of this pulsar in Fig.~\ref{fig:polfractions} reveals that the swing is coincident with a minimum in the linear polarisation fraction, which is characteristic of the presence of OPM jumps (see Section~\ref{subsec:b0950}). Similar behaviour has also been observed by Suleimanova \& Pugachev (2002)\nocite{sp02} at 103 MHz. Hence, if we consider this feature to be an OPM jump, this pulsar's PA profile is rather flat and we cannot constrain the value of $\phi_0$ with an RVM fit.

As was mentioned above, interstellar scattering can cause flattening of the PA profiles and depolarisation of the pulsed emission. In our sample, PSRs B0136+57, B1911$-$04 and B2111+46 are characteristic examples of a pulsar with a scattered profile, the latter of the three having no detectable linear polarisation. For those cases where the effects of scattering appear to dominate over the intrinsic PA evolution across the pulse, it was not possible to obtain a value for $\phi_0$.

Finally, for several pulsars, good examples of which are PSRs B0823+26 and B1919+21, it was deemed necessary to improve the RVM fit by zero-weighting PAs in phase ranges across which the smooth evolution of the PA profile is distorted by local features. Those phase ranges have been greyed out in Fig.~\ref{fig:polprofiles}. In the same figure, we also show all the attempted RVM fits with green lines. The phase at the inflexion point of the RVM curve is marked with a vertical, dotted line. We would like to stress that typically such RVM fits result in large uncertainties on $\alpha$ and $\zeta$, due to the strong co-variance between these parameters that is accentuated by weak or missing polarisation at the edges of profiles. Nevertheless, our primary purpose was to use the RVM model to determine $\phi_0$, which is less sensitive to missing information on the PA.

\subsubsection{Determining $\phi_{\rm em}$}
Secondly, one must determine the phase corresponding to emission generated nearest to the magnetic pole. Typically, this is based on the shape of the pulse profile, where pulse symmetry and component multiplicity are considered. It is difficult to know how the emission is distributed throughout a pulsar's beam, since our line of sight only samples a small cross section of it. Depending on whether the intensity profile corresponds to a cross section of conal or core emission, $\phi_{\rm em}$ can be identified either as the phase at the peak of the profile (e.g.~PSR B1929+10) or the mid-point of a double-peaked profile (e.g.~PSR B1133+16), respectively (Rankin 1983\nocite{ran83}). The double-peaked profiles of PSRs B0809+74, B0834+06 and B1133+16 are normally thought to be conal emission centred on the magnetic pole, so we have chosen $\phi_{\rm em}$ at the profile's midpoint. Although PSRs B0950+08 and B1919+21 also appear as conal doubles, their classification is not as clear: the components of PSR B0950+08 appear clearly distinct at 150 MHz, but their separation decreases with frequency, and the profile nearly becomes a single core component above 1400 MHz; the opposite is true for PSR B1919+21, which appears as a conal double at high frequencies but whose components begin to merge together towards the LOFAR band (see Fig.~\ref{fig:polfractions}). For those two cases, we have also used the profile's midpoint. All the determined values of $\phi_{\rm em}$ are shown in Fig.~\ref{fig:polprofiles} with vertical dashed lines. We stress here that the choice of $\phi_{\rm em}$ is subjective and, apart from the measurement uncertainty, we have no way of quantifying the uncertainty associated with our choice.  

\begin{table*}
\caption{Determined values of the phase lag between the peak (or midpoint) of the pulse profile and the inflexion of the PA profile, $\Delta\phi$, shown in degrees in Col.~2, for 11 pulsars observed at 150 MHz. For four pulsars, for which the peak (or midpoint) of the profile precedes the PA inflexion within the quoted $1\sigma$ uncertainties, Col.~3 shows the emission height, $r_{\rm em}$, calculated in the framework of BCW (Eq.~\ref{eq:bcwlag}). For the rest of the pulsars, $\Delta\phi$ is negative within $1\sigma$ or its sign cannot be confidently determined; therefore, an emission height was not calculated for those pulsars. Col.~4 shows the light-cylinder radius, $R_{\rm LC}$, of each pulsar, in km.} 
\label{tab:emheights}      
\centering
\renewcommand*{\arraystretch}{1.5}
\begin{tabular}{l r r r}        
\hline\hline                 
PSR & $\Delta\phi$ [deg] & $r_{\rm em}$ [km] & $R_{\rm LC}$ [km]   \\    
\hline
B0809+74     & 10$^{+22}_{-23}$             & --                    & 61,657   \\
B0823+26     & 1$^{+1}_{-1}$                & 144$^{+136}_{-134}$   & 25,320   \\
B0834+06     & 2$^{+2}_{-2}$                & 452$^{+437}_{-432}$   & 60,776   \\
B0950+08     & 0$^{+5}_{-5}$                & --                    & 12,075   \\
J1022+1001   & $-$2$^{-2}_{+2}$             & --                    & 785      \\
B1133+16     & 1.4$^{+0.6}_{-0.6}$          & 349$^{+158}_{-150}$   & 56,679   \\
B1508+55     & $-$1.9$^{-0.5}_{+0.5}$       & --                    & 35,293   \\
B1919+21     & 2$^{+4}_{-3}$                & --                    & 63,807   \\
B1929+10     & $-$10$_{-33}^{+11}$          & --                    & 10,808   \\
B1953+50     & 1.6$^{+0.7}_{-1.1}$          & 177$^{+74}_{-119}$    & 24,760   \\
B2224+65     & 2$^{+15}_{-16}$              & --                    & 32,566   \\
\hline
\end{tabular}
\\
\flushleft
\end{table*}

\subsection{Results}
Using the determined phases, we have attempted to estimate the emission heights corresponding to the observed phase lag between $\phi_{\rm em}$ and $\phi_0$ at 150 MHz. All the determined phase lags, $\Delta\phi=\phi_0-\phi_{\rm em}$, and their $1\sigma$ uncertainties are shown in the second column of Table~\ref{tab:emheights}. For four pulsars we found that the BCW condition is satisfied, with $\Delta\phi>0$ within $1\sigma$. For the rest of the pulsars, the RVM fit resulted either in $\Delta\phi<0$ within $1\sigma$ or in a value whose sign could not be confidently determined within the uncertainties. Therefore, for those pulsars we did not calculate an emission height. Finally, the PA profiles of all the remaining pulsars presented in this paper were too flat to provide a constraining fit. 

The emission heights of the four pulsars for which the direction of the phase lag was consistent with BCW were calculated based on Eq.~\ref{eq:bcwlag} and are shown in the third column of Table~\ref{tab:emheights}. The $1\sigma$ uncertainties on $r_{\rm em}$ range from $\approx 50\%$ (for the regularly shaped PA profile of PSR B1133+16) to nearly 100\%, for PSR B0834+06. For three pulsars for which we were able to constrain $\phi_0$ but which did not satisfy the BCW condition, emission heights based on Eq.~\ref{eq:bcwlag} could not be calculated. A number of explanations have been put forward for phase lags in the opposite direction to the model of BCW. For example, it could be that the emission mechanism is not curvature radiation, as is assumed in BCW, but direct or inverse Compton, or even synchrotron emission. Those mechanisms are not likely to be affected by the macroscopic acceleration of co-rotation, since the corresponding microscopic acceleration of the particles is significantly larger (Takata, Chang \& Cheng 2007\nocite{tcc07}). 

We would like to stress again that the presented emission heights are based on the subjective assumption that the maximum or mid-point of the pulse profile corresponds to emission from nearest to the magnetic pole. It is quite possible that our choice of $\phi_{\rm em}$ is erroneous and that those pulsars that appear inconsistent with the model of BCW do in fact also obey their delay--radius relation. Conversely, we cannot exclude the possibility that some or all of the pulsars shown in Table~\ref{tab:emheights} are inconsistent with the BCW model.

Bearing the above caveat in mind, our values of $r_{\rm em}$ can be compared with those found by Hassall et al.~(2012)\nocite{hsh+12} for PSRs B0809+74, B1133+16 and B1919+21. The latter work placed upper limits on the height difference between the radio-emitting regions at 40 and 180 MHz, using data from simultaneous LBA--HBA observations. Those upper limits were estimated from the maximum delay between the time of arrival of the pulses at 180 and 40 MHz, due to aberration/retardation, after having modelled and estimated frequency-dependent delays caused by the ISM. In addition, under the assumption of RFM, the authors were able to provide an upper limit on the absolute height of the lowest-frequency emission observed, i.e.~40 MHz. This was done by combining the upper limit on the delay due to aberration/retardation with the pulse broadening measured across the investigated frequency range. Of the three pulsars mentioned above, only the profile evolution of PSR B1133+16 agreed with RFM and was therefore the only pulsar for which an upper limit on the absolute height could be placed. The published upper limit from that work for PSR B1133+16 is $110$ km. We also note that that the previous estimate by Kramer et al.~(1997)\nocite{kxj+97}, who performed a similar analysis to Hassall et al., but using only high frequencies, yielded a less constraining upper limit of 320 km. 
   
In our work, the large uncertainties on the determined $\phi_0$ for PSRs B0809+74 and B1919+21 did not allow us to constrain the emission height. In contrast, the value of $\phi_0$ for PSR B1133+16 is fairly constrained from the RVM fit. This pulsar's profile is roughly symmetric with respect to the PA inflexion, which suggests that its two brightest components are likely generated on opposite sides of the fiducial plane containing the the spin and magnetic axes. Hence, for PSR B1133+16 we have chosen $\phi_{\rm em}$ to be the mid-point between the two brightest components. This choice yields an emission height of $r_{\rm em}=349^{+158}_{-150}$ km, the $1\sigma$ interval of which is 1.5--4.5 times larger than the upper limit of Hassall et al.~(2012).

The reason for the inconsistency between the emission height based on polarisation and that based on pulsar timing could be that our choice of the fiducial points in the profile of PSR B1133+16 deviate from the actual ones. As was mentioned above, the choice of $\phi_{\rm em}$ is only based on the observed profile's mid-point between the maxima of the two main components. A different definition of the mid-point, for example by weighting the position of the fiducial point based on the integrated flux of each component, could easily shift $\phi_{\rm em}$ to later phases, which would lower the emission height accordingly. Conversely, it is not uncommon in pulsar-timing measurements that the associated uncertainties are underestimated. This could be especially true in a low-S/N environment such as the LBA measurements of Hassall et al.~(2012). In such case, an underestimation of the timing uncertainties by less than a factor of two would proportionally result in the underestimation of the emission height by the same amount, in that work; and this would make our measurements consistent with the upper limit from pulsar timing. Most likely, a combination of under-/overestimation of the derived emission heights from both methods could be the explanation for the observed inconsistencies.

In summary, our emission-height estimates for PSRs B0823+26, B0834+06, B1133+16 and B1953+50 are all consistent with the emission region located a few hundred km above the pulsar surface. For PSR B1133+16, our polarisation measurements in combination with the delay--radius relation of BCW yield an emission height that is up to a few times larger than the upper limit from pulsar timing. However, it is likely that the values derived from either method are subject to mostly unquantifiable systematic uncertainties.

\section{Individual pulsars}
\subsection{PSR B0950+08}
\label{subsec:b0950}
PSR B0950+08 is an interesting special case, since it exhibits significant polarisation-profile evolution between 150 MHz and 1400 MHz. At 150 MHz, the profile comprises two distinct, highly polarised components that begin to merge together towards higher frequencies, while at the same time becoming more weakly polarised (see Fig.~\ref{fig:polfractions}; components L and T). The depolarisation of PSR B0950+08 due to overlapping modes was also noted by Gangadhara (1997)\nocite{gan97}. This is expected in the framework of birefringence. However, the PA profile at 150 MHz clearly shows that both the leading and trailing component emit in the same polarisation mode (i.e.~there are no evident orthogonal jumps). Hence, if the two components emit in the same mode between 150 and 1400 MHz, we do not expect depolarisation. However, at 600 and 1400 MHz, the PA profile develops a bump at $\phi\approx 0.51$, with its leading edge having $|\Delta {\rm PA}|\approx 70^\circ$ and its trailing edge, $|\Delta {\rm PA}|\approx 90^\circ$. This feature is $\approx 1.5$ times broader at 1400 MHz than at 600 MHz. The appearance of the bump at 600 MHz is accompanied by that of a third component in the linear polarisation profile (component C in Fig.~\ref{fig:polfractions}), which is coincident with the bump and resides between the two components that are present up to 600 MHz. Evidently, component C emits in the orthogonal mode to its neighbouring components, which would cause depolarisation at the overlapping regions with those. Indeed, at the phases where the orthogonal jumps occur in the 600 MHz profile, the linear polarisation dips. Furthermore, at 1400 MHz component L becomes very weak, and only components C and T are clearly visible. In addition, it can be seen that the width of component C follows the frequency evolution of the bump in the PA profile. It is unclear whether component C is a completely independent magnetospheric emission region, only seen at high frequencies, or a by-product of the interference between the other two components as they merge together. Certainly the intensity and width of this central component correlates well with the degree of mixing between the other two components --- although it must be noted that the overall pulse width, and hence the component separation between 600 MHz and 1400 MHz remains roughly the same.

\begin{figure}
\centering
\includegraphics[height=\dimexpr \textheight - 10\baselineskip\relax,width=0.47\textwidth,keepaspectratio]{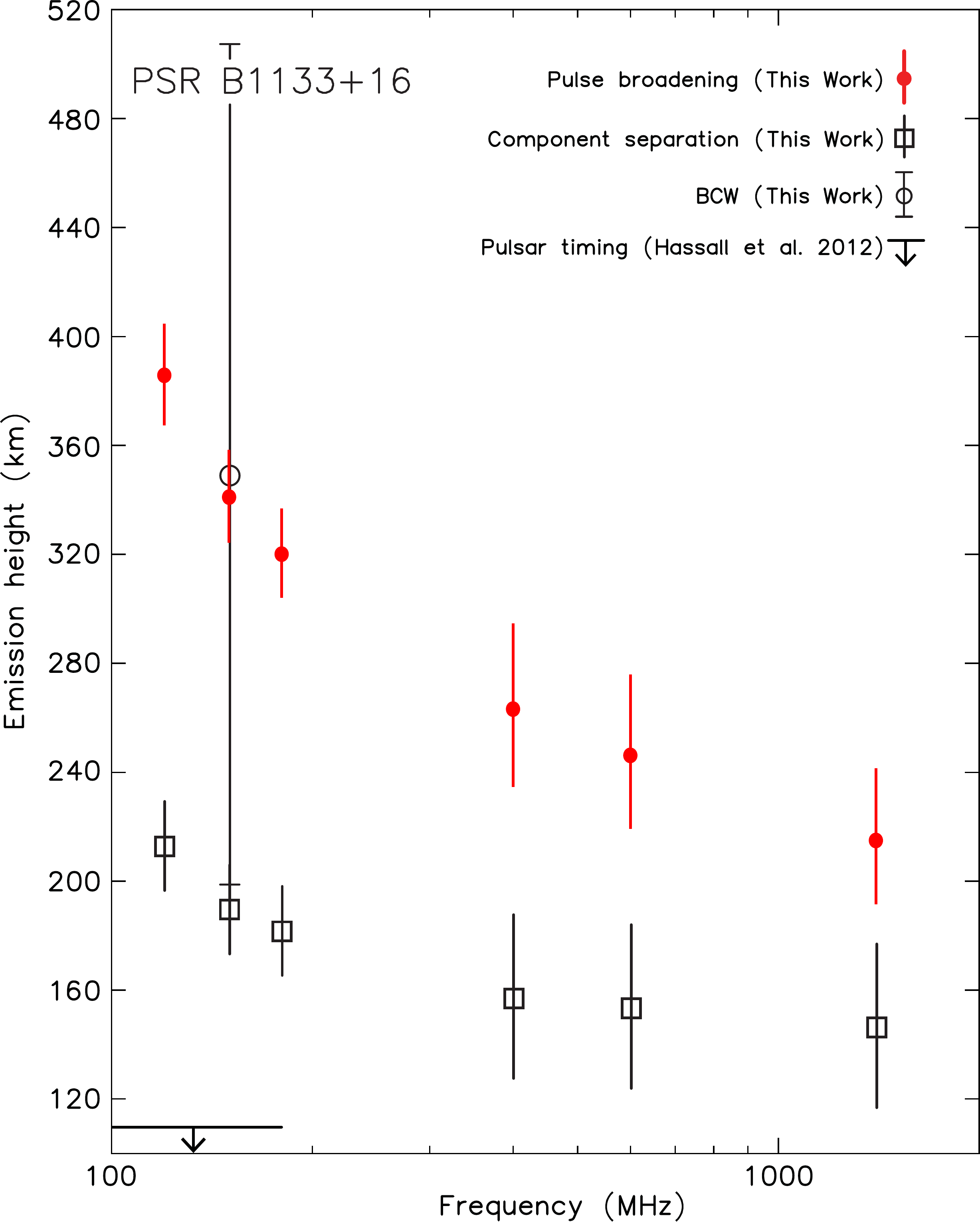}
\caption{Emission heights at different observing frequencies, for PSR B1133+16, based on pulse broadening (filled red circles) and component separation (open squares), and the assumption that the observed emission is coupled to the last open field-lines of the pulsar's dipolar magnetic field. For comparison, we also show with an open circle the emission height estimate based on the delay--radius relation of Blaskiewicz, Cordes and Wasserman and our polarisation observations at 150 MHz (Table~\ref{tab:emheights}). Finally, the upper limit on the emission height from pulsar-timing measurements by Hassall et al.~(2012), between 40 and 180 MHz, is shown with an arrow.}
\label{fig:b1133heights}%
\end{figure}

\subsection{PSR B1133+16}
\label{subsec:b1133}
PSR B1133+16 has a double-peaked profile, where its components become clearly more separated towards low frequencies. This was one of the pulsars that motivated RFM (Komesaroff 1970\nocite{kom70}), under the assumption that the emission corresponding to the two components is coupled to the last open field-lines. In addition, the PA profile of this pulsar seems to follow the regular swing that is expected by the RVM, very well. Therefore, it is justified to attempt and combine the viewing geometry of this pulsar, derived from polarisation, with its pulse broadening, and estimate the emission height as a function of frequency. Gil et al.~(1984)\nocite{ggr84} provide an expression, based on the curvature of the dipolar-field lines, that allows the calculation of the angular radius of the emission cone, $\rho$, as a function of the pulse width and the $\alpha$ and $\zeta$ parameters. The values of $\alpha$ and $\zeta$ for this pulsar have been determined via RVM fits at 400 MHz, by Lyne \& Manchester (1988)\nocite{lm88}, i.e.~$\alpha=51.3^\circ$ and $\zeta=55^\circ$. Furthermore, under the assumption that at all frequencies the emission is coupled to the last open field-lines, one can relate $\rho$ to the emission height, i.e.~$\rho\approx 86^\circ(r_{\rm em}/R_{\rm LC})^{1/2}$ (see e.g.~Gangadhara \& Gupta 2001\nocite{gg01}). We should note that this expression implicitly assumes a perfectly aligned rotator, where the maximum co-rotating radius along the last open field line is $R_{\rm LC}$. In reality, for arbitrary values of $\alpha$, larger co-rotating radii are allowed and the expression becomes more complex (Lee et al.~2009\nocite{lcw+09}). We have used the pulse widths derived for this pulsar in Section~\ref{subsec:polfracs} to calculate the emission height as a function of observing frequency. The emission height at each frequency band, based on pulse broadening, is shown in Fig.~\ref{fig:b1133heights}. As expected from assuming that the emission is bounded by the open field-lines, the height increases roughly two-fold between 1400 MHz and 120 MHz. It can also be seen that below 200 MHz the upper limit from pulsar timing constrains the height to approximately half the value that is derived from pulse broadening. 

Our definition of the pulse width yields larger values than the phase separation between the maxima of the two brightest components, in the total intensity profile. If the separation between the components at different frequencies is used instead, then, as Fig.~\ref{fig:b1133heights} shows, the evolution of the emission height between 1400 MHz and 120 MHz is greatly attenuated. More specifically, using our definition of the pulse width results in a height differential of $\Delta r_{\rm em}\approx 150$ km, between 1400 MHz and 100 MHz. On the other hand, using the component separation as our prior results in only $\Delta r_{\rm em}\approx 50$ km; in particular, between 1400 MHz and 400 MHz the emission height remains roughly constant ($r_{\rm em}\approx 150$ km), within the uncertainties. 

Compared to the upper limit by Hassall et al.~(2012), the values based on pulse broadening and component separation yield larger values by a factor of a few. This inconsistency could be considered further evidence against RFM, at least for this pulsar. It is quite possible that the pulse broadening we observe is only partly or even not at all due to RFM. For example, in the study of McKinnon et al.~(1997)\nocite{mck97}, the mechanism of birefringence was put forward as an argument for pulse broadening towards lower frequencies: this mechanism is independent of RFM yet acts in the same direction to cause pulse broadening. In such case, if only part of the broadening is due to RFM, the calculated emission heights from pulse broadening and component separation should be considered as upper limits. However, it is important to note that this pulsar's PA profile below 1400 MHz is devoid of OPM jumps. This fact may advocate against pulse broadening being the result of OPM separation towards lower frequencies, as one would expect if birefringence was in play.  

Finally, as was noted earlier, the above calculations assume that at all frequencies the emission is coupled to the last open field-lines, which may not be true. If indeed the different frequencies come from different magnetic field lines within the open field-line region, then it is possible that RFM is invalid and that all emission originates from roughly the same height --- or even that high-frequency radiation is generated higher in the magnetosphere than low-frequency radiation.

\subsection{Millisecond pulsars}
\label{subsec:msps}
In addition to the non-recycled pulsars, our observations included four MSPs, PSRs J0034$-$0534, J1012+5307, J1022+1001 and B1257+12. In general, MSPs are thought to be old, recycled pulsars, with characteristic ages of several hundreds of Myr. The dipolar magnetic fields of MSPs are three to four orders of magnitude weaker than those of non-recycled pulsars and they are also confined within a much smaller light cylinder, since $R_{\rm LC}\propto P$. This results in wider solid angles of emission, as defined by the open field-line region above the polar caps, which is observationally supported by the larger pulse duty cycles of MSP profiles compared to those of non-recycled pulsars. As concerns the polarisation properties of MSPs, they can also exhibit high degrees of linear polarisation and other polarisation features (e.g.~orthogonal jumps) found in non-recycled pulsars' profiles. However, systematic studies of MSP polarisation have revealed that they possess much more complex PA profiles than those of non-recycled pulsars (Yan et al.~2011\nocite{ymv+11}).

We will now discuss individually the polarisation properties of each MSP:

\subsubsection{PSR J0034$-$0534}
\label{subsubsec:mspj0034}
This binary MSP was discovered by Bailes et al.~(1994)\nocite{bhl+94} in a survey of the southern sky with Parkes. It is a relatively nearby pulsar: according to its low DM of $\approx 13.8$ pc cm$^{-3}$ (Abdo et al.~2010\nocite{aaa+10b}) and the NE2001 electron-density model, its estimated distance is $d=0.53$ kpc (Cordes \& Lazio 2002\nocite{cl02}). The only polarisation measurements of this pulsar to date were performed at 400 MHz with the Lovell telescope, by Stairs et al.~(1999)\nocite{stc99}. Unfortunately, those observations showed the absence of linear polarisation ($<5\%$) and only a small amount of circularly polarised flux ($\approx 18\%$). Hence, a measurement of the RM for this pulsar has not been obtained. 

The relative proximity of this pulsar combined with its high Galactic latitude ($b\approx -68^\circ$) is expected to result in a small value of RM. We can obtain a rough estimate of the amount of Faraday rotation expected towards PSR J0034$-$0534, based on the amount of Faraday rotation measured for the nearby PSR B0031$-$07: the latter pulsar is only $1.8^\circ$ away in the sky and has a similar but lower DM of $\approx 11.4$ pc cm$^{-3}$ (Hobbs et al.~2004\nocite{hlk+04}). In addition, there is a precise VLBI parallax measurement for PSR J0034$-$0721, which gives a distance of $d=1.06^{+0.08}_{-0.09}$ kpc (Chatterjee et al.~2009\nocite{cbv+09}). Given that neither of those two pulsars has been associated with a dense ISM environment, such as a supernova remnant, and that it is unlikely that the ISM density fluctuates significantly at such high latitudes, the NE2001 distance for PSR J0034$-$0534 seems to be an underestimate. For example, assuming a model of the free-electron density that is exponentially decreasing with Galactic height, $z=\sin b$, and which has a scale height of $H_0\approx1.7$ kpc (Schnitzeler 2012\nocite{sch12}), the difference in DM between the pulsars yields a distance of $1.4$ kpc for PSR J0034$-$0534. 
Under the assumption that the magnetic field between those pulsars remains roughly constant, we expect the RM of PSR J0034$-$0534 to be proportionally higher than that of PSR B0031$-$07 by the amount of additional dispersion. If we take our RM measurement of ${\rm RM}\approx 10$ rad m$^{-2}$ for PSR B0031$-$07, we find that under this assumption the expected RM for PSR J0034$-$0534 is roughly 12 rad m$^{-2}$.

\begin{figure}
\centering
\includegraphics[height=\dimexpr \textheight - 10\baselineskip\relax,width=0.47\textwidth,keepaspectratio]{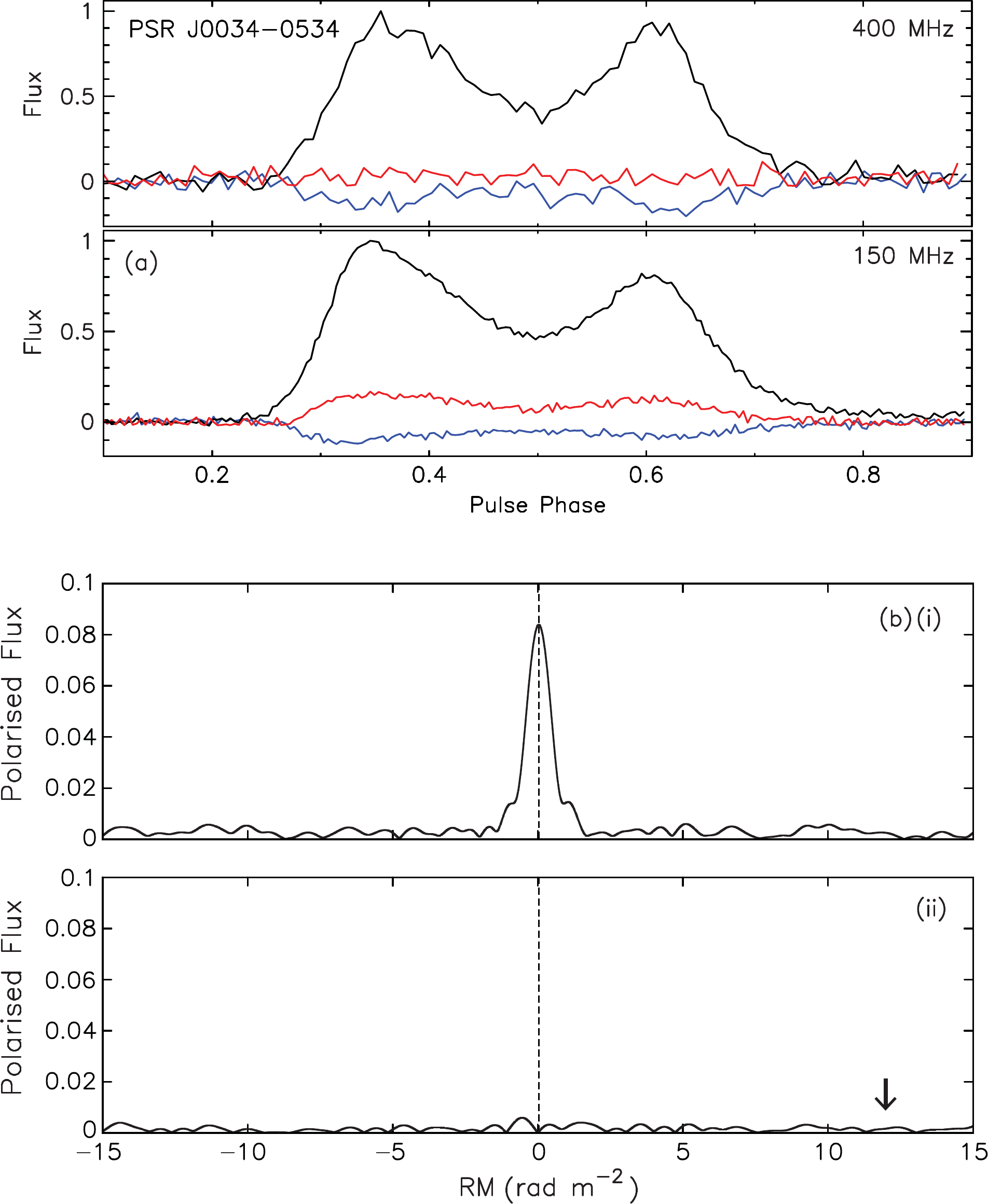}
\caption{(a) Polarisation profile of the MSP J0034$-$0534 at 150 MHz and 400 MHz. All the flux scales are in arbitrary units. (b) RM spectra for PSR J0034$-$0534 from the LOFAR data: (i) RM spectrum before removing the instrumental response, centred at ${\rm RM}\approx0$ rad m$^{-2}$; (ii) spectrum after subtracting the instrumental response from the data. The black arrow indicates the position of the expected RM for this pulsar, under the assumptions stated in Section~\ref{subsubsec:mspj0034}.}
\label{fig:msppol0034}%
\end{figure}

At 150 MHz, our observations show that the small fraction of circular polarisation is maintained ($\approx 12\%$; see Fig.~\ref{fig:msppol0034}a). Before we can determine the amount of linear polarisation, we need to correct for the Faraday rotation. The RM spectrum, shown in Fig.~\ref{fig:msppol0034}b(i) was calculated using the entire available HBA bandwidth of 96 MHz. It can be clearly seen that the peak at ${\rm RM=0}$ rad m$^{-2}$ dominates over every other spectral feature. This peak, as was explained in the introduction, is the result of imperfect instrumental calibration and does not reflect any astrophysical effect. 

The instrumental contribution to the RM spectrum is a well-defined sinc function, given the frequency coverage at a given observing frequency. Therefore, it can be subtracted from the spectrum, which is to a first order the application of the \textsf{RM-CLEAN} process (Heald et al.~2009\nocite{hbe09}). After cleaning the original spectrum, we are left with the spectrum shown in Fig.~\ref{fig:msppol0034}b(ii). The cleaned spectrum does not contain any prominent peaks, neither near the expected RM value from our comparison with PSR B0031$-$07, nor elsewhere inside the wide range of RM investigated. Hence, we are forced to conclude that we have not been able to detect any significant linear polarisation of astrophysical origin, for PSR J0034$-$0534. In the future, it may be possible to perform more sensitive observations and a better, more precise polarisation calibration for this pulsar, and measure its RM; but this is beyond the scope of the current paper.

\subsubsection{PSR J1012+5307}
\begin{figure*}
\centering
\includegraphics[height=\dimexpr \textheight - 10\baselineskip\relax,width=\textwidth,keepaspectratio]{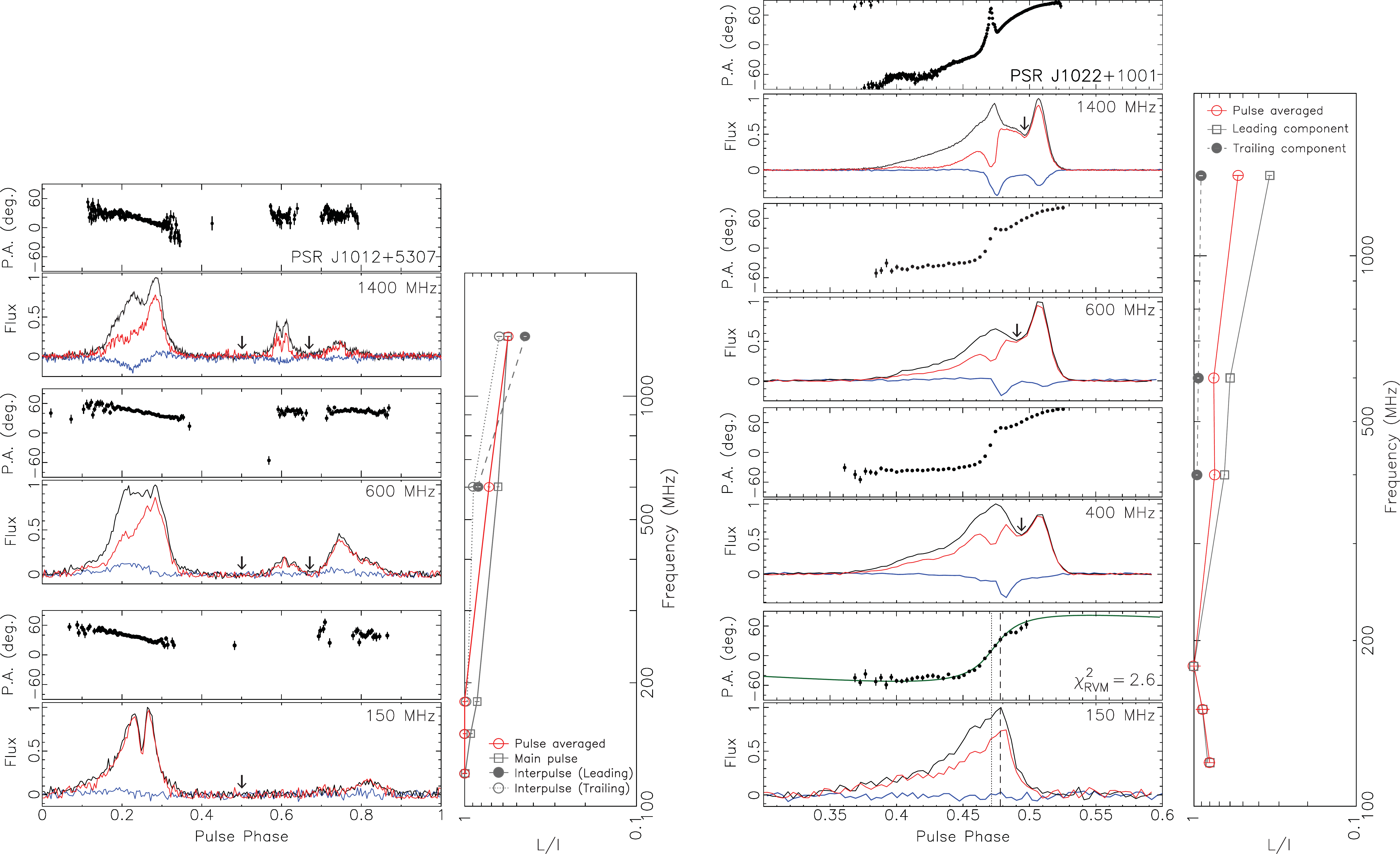}
\caption{Evolution of the polarisation profiles as a function of observing frequency, for PSRs J1012+5307 and J1022+1001 in our sample. All the flux scales are in arbitrary units. The high-resolution ($\sim 5$ $\upmu$s) profiles at 1400 MHz were obtained by Xilouris et al.~(1998) for PSR J1012+5307, and Yan et al.~(2011) for PSR J1022+1001. Alongside the profiles, the fraction of linear polarisation as a function of observing frequency is also shown for the pulse averaged flux (red open circles). The linear polarisation fraction as a function of frequency for the phase ranges demarcated by the black arrows is shown with grey symbols. For PSR J1021+5307, the linear polarisation fraction of the main pulse is shown with grey squares; that of the leading and trailing component of the interpulse is shown with grey filled circles and a grey open circles, respectively. For PSR J1022+1001, the linear polarisation fraction for the leading and trailing components are shown with grey squares and grey filled circles, respectively. The best RVM fit to the 150 MHz PA profile of PSR J1022+1001 is show with a green line. The determined value of $\phi_0$ from the RVM fit and the phase at the profile's maximum are shown with vertical dotted and dashed lines, respectively.}
\label{fig:msppaprofs}%
\end{figure*}

Discovered during a survey for short-period pulsars with the 76 m Lovell telescope, PSR J1012+5307 is a 5.3 ms binary pulsar with a white dwarf companion. The polarisation of this pulsar has been studied at 600 MHz and 1400 MHz (Xilouris et al.~1998\nocite{xkj+98}; Stairs et al.~1999\nocite{stc99}). The corresponding profiles are shown in Fig.~\ref{fig:msppaprofs}. In general, the profile of this pulsar at those frequencies is complex, composed of a main pulse (MP) and an interpulse (IP), roughly 180$^\circ$ away. Both MP and IP are composed of at least two components, with the leading component of the IP being divided into two components at 1400 MHz but being completely absent at 150 MHz. The MP shows an interesting evolution between 1400 MHz and 600 MHz. Observations with the Green Bank Telescope at 1400 and 800 MHz (not shown here) have revealed that at those frequencies this pulsar's MP may be composed of up to six components (Dyks, Wright \& Demorest 2010\nocite{dwd10}). At 150 MHz, two components of the MP are clearly visible, of roughly equal magnitude and much more clearly separated than at higher frequencies. 

In terms of polarisation, all visible components remain highly linearly polarised across all investigated frequencies, with the LOFAR profile being $\approx 100\%$ linearly polarised. We have performed a simplified component-by-component analysis of the fraction of linear polarisation (see Fig.~\ref{fig:msppaprofs}), where we have only considered the phase windows including the MP and the two components of the IP. Our analysis shows that the linear polarisation fraction increases monotonically with decreasing observing frequency for all components. As was noted in Xilouris et al.~(1998)\nocite{xkj+98} and Stairs et al.~(1999)\nocite{stc99}, the 1400 MHz and 600 MHz profiles are moderately circularly polarised, having fractions of $\approx 17\%$ and $\approx 10\%$, respectively. At 150 MHz, the circular polarisation fraction remains low, at $\approx 9\%$. 

The PA profile of PSR J1012+5307 at 150 MHz exhibits a small slope across the MP and is practically flat across the IP (see Fig.~\ref{fig:msppaprofs}). This is similar to what is observed at the higher frequencies. Interestingly, the PA slope across the MP at 600 MHz is $\approx 20\%$ flatter than both the 1400 MHz and 150 MHz profiles. This could be related to the fact that the total flux density of the MP resembles a top-hat function, with its constituent components appearing merged together. This could possibly lead to depolarisation and flattening of the PAs, similarly to scattering (see Section~\ref{sec:rmvars}). Finally, due to the flatness of the PA profile, across both the main pulse and the interpulse, our attempt to fit an RVM to the PAs of this MSP resulted in an unconstrained geometry.

\subsubsection{PSR J1022+1001}
\label{subsubsec:mspj1022}
PSR J1022+1001 is a 16 ms binary pulsar discovered at 430 MHz, with the Arecibo telescope (Camilo et al.~1996). Its total intensity profile displays a complex frequency evolution, which has been studied with Effelsberg and WSRT observations, between 328 MHz and 4.8 GHz, by Kramer et al.~1999\nocite{kll+99} and Ramachandran \& Kramer (2003)\nocite{rk03}, respectively. The average pulse profile of PSR J1022+1001 is composed of two components with different spectral characteristics: in the investigated range, the leading component is dominant at the highest ($>1400$ MHz) and lowest ($<350$ MHz) frequencies, while at $\approx 400$ MHz and $\approx 1400$ MHz the two components have equal strength, and finally between 500 MHz and 1 GHz, the trailing component dominates. In addition, the polarisation properties of this pulsar display a complex behaviour. The PA profile between 400 MHz and 1400 MHz resembles a typical RVM swing but for two distorting features. Firstly, it exhibits a notch that is roughly coincident with the maximum value of $|V|$ (see Fig.~\ref{fig:msppaprofs}; $\phi\approx 0.47$). Secondly, as was noted by Ramachandran \& Kramer, the leading part of the PA profile at 1400 MHz exhibits a bump that appears coincident with a weak leading component in the linear polarisation profile ($\phi\approx 0.40$). Both of these features are evident in the high-resolution profile obtained at 1400 MHz with the Parkes telescope by Yan et al.~(2011)\nocite{ymv+11}. The origin of these distorting features has been investigated by Ramachandran \& Kramer (2003)\nocite{rk03}, who suggested that they can be explained by the presence of magnetospheric return currents that are subjected to aberration due to the larger emission heights involved in MSPs (Kramer et al.~1998\nocite{kxl+98}).

At 150 MHz, the profile of PSR J1022+1001 contains only a single component with a long leading tail, which is also seen to precede the leading of the two principal components at higher frequencies (see Fig.~\ref{fig:msppaprofs}). By aligning the PA profiles between 1400 MHz and 150 MHz to the phase at the steepest PA gradient, we can identify the emission seen at 150 MHz as that corresponding to the leading component seen at higher frequencies (see Fig.~\ref{fig:msppaprofs}). The phase at the steepest PA gradient at 150 MHz was determined by fitting an RVM model to the data ($\phi_0\approx 0.471$). If the above is true, we must conclude that the trailing component vanishes somewhere between 320 MHz (the bottom of the WSRT band) and 200 MHz (the top of the LOFAR 150 MHz band). As was noted in previous studies, the principal components of the pulse profile have significantly different polarisation fractions at 1400 MHz, with the bridge emission and the trailing component being nearly 100\% polarised ($\phi>0.48$), while the leading component and tail ($\phi<0.48$) are $<50\%$ polarised. In addition both components are seen to have significant circular polarisation, which is maintained down to 400 MHz for the leading component but disappears almost completely at that frequency for the trailing component. After correcting for Faraday rotation (${\rm RM}=2.18(2)$ rad m$^{-2}$), at 150 MHz the single component visible is highly linearly polarised, with a pulse-averaged polarisation of $\approx 80\%$. In addition, the circular polarisation vanishes completely at LOFAR frequencies. A comparison of the linear polarisation fraction between 1400 MHz and 150 MHz shows that there is a 45\% increase. Interestingly, within the LOFAR band we measured a small decrease in the linearly polarised fraction below 180 MHz, perhaps indicating a turnover (see Fig.~\ref{fig:msppaprofs}).

Furthermore, we find that at 150 MHz the PA profile of PSR J1022+1001 is devoid of the notch feature seen at 1400 MHz, and which is seen to gradually disappear already below 600 MHz (Fig.~\ref{fig:msppaprofs}). However, there is a hint of the leading bump at $\phi\approx 0.42$, which is actually not present in the 400 and 600 MHz profiles. Altitude-dependent polarisation effects have been proposed to explain this pulsar's distorted PA profile. Preliminary fits to the Hibschman \& Arons (2001)\nocite{ha01} model suggest an emission height of $0.4R_{\rm LC}\approx 800$ km (Ramachandran \& Kramer 2003\nocite{rk03}). Based on our RVM fit, at 150 MHz the peak of the emission lags the phase at the PA inflexion by $\Delta\phi=2^\circ(1)$. Like in the case of PSR J1012+5307, the direction of the lag contradicts the BCW delay--radius relation. 

\section{Summary and conclusions}
\label{sec:summary}
We have undertaken a detailed investigation of the polarisation properties of 20 bright pulsars, between 105 and 197 MHz, using the LOFAR HBA core. This is the first time high-quality polarisation data are obtained at these frequencies, for this sample of pulsars. The high frequency and time resolution, as well as the high fractional bandwidth available with LOFAR, have allowed us to measure polarisation fractions at LOFAR frequencies, with high precision. Subsequently, by combining these measurements with those previously published at 1400, 600, 240 and 400 MHz, we have measured the spectrum of fractional polarisation between 1400 MHz and 100 MHz for those pulsars. Interestingly, after excluding pulsars that are subjected to strong interstellar scattering that possibly leads to depolarisation towards low frequencies, we could only find one pulsar in our sample, PSR B1508+55, whose fractional polarisation increases with observing frequency. On the other hand, we find six pulsars for which polarisation decreases with frequency. In some cases we note a turn-over in the spectrum of fractional polarisation, within the investigated range. For some of the pulsars, like PSR B1911$-$04, such a turn-over could be the result of depolarisation due to scattering, combined with the intrinsic spectrum of polarised emission: e.g.~the frequency dependence of depolarisation could make it more dominant than the spectrum at lower frequencies, whereas the opposite may true at higher frequencies. 

Furthermore, we have tested the predictions of birefringence in pulsar magnetospheres by investigating a possible anti-correlation between the pulse width and the polarisation fraction as a function of observing frequency. This is an important test, as the mechanism of birefringence could explain e.g.~orthogonal polarisation modes and intrinsic pulse broadening at low frequencies. Unfortunately, we could not find strong evidence for such a mechanism via this process, with only 60\% of our sample satisfying such an anti-correlation. 

Beyond the effects of pulsar magnetospheric emission, we have also investigated the effects of the interstellar medium on polarisation. It has been previously reported, based on 1400 MHz data, that interstellar scattering coupled with steep PA profiles causes an apparent variation of the amount of Faraday rotation as a function of pulse phase. In this paper, we investigated (a) how evident this effect is at low radio frequencies and (b) how the magnitude of this effect scales between the 1400 MHz and 150 MHz. To this aim, we have measured the amount of Faraday rotation as a function of pulse phase for eight pulsars at 150 MHz. Interestingly, we have found that the typical magnitude of the variations at 150 MHz is $\sim 100$ times lower than what has been measured at 1400 MHz, for a different sample of pulsars. We have used a simple model to investigate whether the observed effects could be caused by scattering. We have found that indeed scattering can introduce changes in the PA with frequency that would mimic phase-dependent Faraday rotation. Moreover, it is predicted that the maximum variation of the Faraday rotation introduced by scattering, across the profile, should scale with wavelength as $1/\lambda^2$. The two orders of magnitude difference in the typical magnitude of the variations between 1400 MHz and 150 MHz is consistent with the model's prediction, which provides further support for scattering being the source of the observed variations.

The high-S/N profiles have also allowed us to estimate the magnetospheric height of the 150 MHz emission, based on the delay--radius relation of Blaskiewicz, Cordes \& Wasserman and the lag between the phase at the steepest PA gradient and that at the profile's maximum intensity (or mid-point, depending on the profile's complexity). Using the observed phase lags, we have estimated emission heights for four pulsars, for which the relation of BCW is applicable. 

For all pulsars for which we were able to constrain the emission height, our polarisation measurements are consistent with the 150 MHz emission being generated at heights of a few hundred km above the pulsar surface. 
For PSR B1133+16, in particular, our estimate of the emission height is larger by a factor of a few than the upper limit of Hassall et al.~(2012) and consistent within the uncertainties with the height obtained from pulse broadening and assuming radius-to-frequency mapping. Nevertheless, due to unquantifiable systematic uncertainties in the determination of the fiducial phase of pulsar emission, we cannot make a conclusive statement as to whether our findings suggest that radius-to-frequency mapping is valid or not. We hope that in the future an ensemble of emission heights based on pulsar timing at low frequencies will be available for comparison against an equally large sample of emission heights based on polarisation. A comparison of those two samples with the predictions of radius-to-frequency mapping could be the first step towards a conclusive statement about the validity of the latter.

Finally, the four MSPs in our sample were discussed separately from the rest of the sample. PSR B1257+12 was left out of this discussion, as polarisation profiles at higher frequencies are either not available or of very low quality. PSR J0034$-$0534 has not had an RM measurement to date, due to the lack of detectable linear polarisation above 400 MHz. Despite the small degree of polarisation present at 150 MHz, it was shown that this is mainly due to instrumental leakage and cannot be disambiguated from the real polarisation, if any exists. Future improvements in LOFAR calibration may be able to remove any instrumental effects and detect any weak polarisation from this pulsar. PSR J1012+5307 has a complex profile, composed of a highly linearly polarised pulse and interpulse. The polarisation of this pulsar increases monotonically towards low frequencies, while notably the leading component of the interpulse vanishes below 400 MHz. 
The final MSP, PSR J1022+1001, exhibits an intriguing frequency evolution from 1400 MHz down to 150 MHz where the trailing component of the profile vanishes along with any detectable circular polarisation. Interestingly, the pulse-averaged linear-polarisation fraction marginally increases between 400 MHz and 180 MHz but seems to turn over below that frequency. PSR J1022+1001 is one of only four pulsars that show such a turn-over
in the LOFAR band. Two of those four pulsars possess long scattering
tails at 150 MHz, which may cause such a turn-over through
depolarisation. Consequently, we show for the first time an intrinsic
turn-over in the polarisation of two pulsars, namely PSRs J1022+1001
and B1237+25.

In summary, our work has highlighted the importance of low-frequency polarisation in the efforts of understanding the elusive magnetospheric processes that lead to pulsar radio emission. In those efforts, a better understanding of how the ISM distorts the intrinsic pulsar emission is pivotal. At the same time, low frequencies provide an excellent opportunity for studying the ISM, where its effects become more pronounced. LOFAR has the sensitivity at low frequencies that can make such studies conclusive. Our work has focused on the complementarity between low-frequency and high-frequency observations of pulsars. Such multi-frequency data have revealed that the evolution of pulsar polarisation cannot be explained solely by means of a single physical process, like birefringence, or the model of radius-to-frequency mapping. On the contrary, it is quite likely that a combination of magnetospheric refraction, occurring over different path lengths at different frequencies, and co-rotational effects, distorting the observed signal, coupled with the intrinsic pulsar spectrum, is what we observe for each pulsar; the particularities of each process are also likely to vary between pulsars. Finally, it is clear that these effects can be masked by scattering, to a different degree for each pulsar. Currently, there are ongoing efforts to map the properties of the ISM via long-term monitoring of pulsar DMs and RMs, using LOFAR. In the near future, using ultra-broadband receivers and ultimately with the advent of the SKA, it will be possible to monitor scattering and dispersion towards a pulsar and try and recover the intrinsic signals. Even before the SKA, it may be possible to simulate the effects of scattering on polarisation --- with more complex models of the scattering screens than what has been assumed in our work --- and perform multi-parametric fits between the observed and simulated polarisation profiles. Ultimately, such fits may yield a large sample of scattering timescales, which can in turn be used towards mapping the small-scale structure of the Galactic ISM.

\begin{acknowledgements}
The authors would like to thank Dr.~Jaros\l{}aw Dyks for a very constructive referee report. AN would also like to thank Dr.~Norbert Wex for invaluable discussions and suggestions on the physics and analytic calculations presented in this paper. In addition, AN would like to extend his thanks to Prof.~Joanna Rankin for stimulating discussions on the polarisation properties of pulsars, during the {\em Extreme Astrophysics in an Ever-Changing Universe} meeting in Ierapetra, Crete.

LOFAR, the Low Frequency Array designed and constructed by ASTRON, has facilities in several countries, that are owned by various parties (each with their own funding sources), and that are collectively operated by the International LOFAR Telescope (ILT) foundation under a joint scientific policy.

J.W.T.H.~acknowledges funding from an NWO Vidi fellowship and ERC Starting Grant ``DRAGNET'' (337062).

MS acknowledges support by the South Africa National Research Foundation Square Kilometre Array Research Fellowship.
\end{acknowledgements}

\bibliographystyle{aa}
\bibliography{journals,modrefs,psrrefs,crossrefs}

\appendix

\section{}
\begin{figure}
\centering
\includegraphics[height=\dimexpr \textheight - 10\baselineskip\relax,width=0.47\textwidth,keepaspectratio]{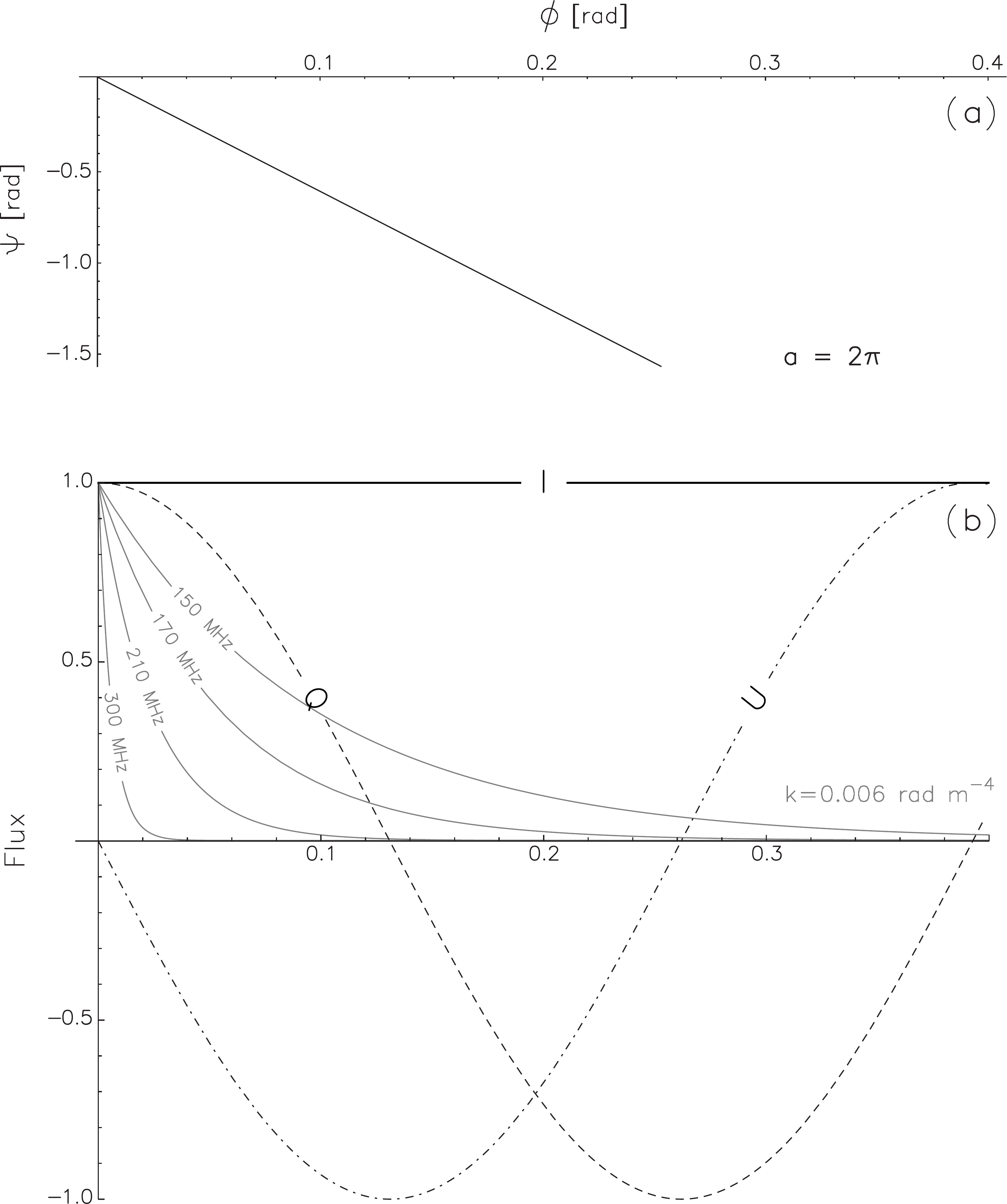}
\caption{(b) The Stokes I, Q and U profiles, and (a) the intrinsic PA profile, $\Psi$. For the example presented here, we have assumed an intrinsic PA gradient $a=2\pi$.}
\label{fig:app1}%
\end{figure}

We consider the simple case of a polarised square pulse of unit amplitude and with a constant PA gradient, $a$ (see Fig.~\ref{fig:app1}). In such case, the Stokes parameters are given by
\begin{equation}
\label{eq:app1}
q=\cos(2\psi) \\
u=\sin(2\psi)
\end{equation}
where $\psi(\phi)=a\phi$ is the PA, and $\phi$ is the pulse phase.

We now assume that the square pulse is scattered by a thin screen and that the scattered intensity has a Gaussian angular distribution; for this example we ignore Faraday rotation and intrinsic profile evolution with frequency. At a given pulse phase, $\phi$, and wavelength $\lambda$, the observed polarisation vector, $\tilde{P}(\phi,\lambda)=Q(\phi,\lambda)+iU(\phi,\lambda)$, is the result of the convolution of the intrinsic polarisation vector, $\tilde{p}(\phi)=q(\phi)+iu(\phi)$, with a one-sided exponential function, ${\rm exp}(-\phi/\tau_{\rm s})$; $\tau_{\rm s}=k\lambda^4$, where $k$ is a constant: 
\begin{equation}
\label{eq:app2}
\tilde{P}(\phi,\lambda)=Q(\phi,\lambda)+iU(\phi,\lambda)=\frac{1}{\tau_{\rm s}}\int_{0}^\phi\tilde{p}(\phi^\prime){\rm e}^{-(\phi-\phi^\prime)/\tau_{\rm s}}{\rm d}\phi^\prime
\end{equation}
The scattered PA profile, $\Psi(\phi,\lambda)$, is then given by
\begin{equation}
\label{eq:app3}
\Psi(\phi,\lambda)=\frac{1}{2}\tan^{-1}\left(\frac{U}{Q}\right)
\end{equation}
Since we are mainly interested in how scattering can have an effect on the measured Faraday rotation, we would like to calculate the derivative of $\Psi$ with respect to $\lambda^2$:
\begin{equation}
\label{eq:app4}
\frac{\partial\Psi}{\partial(\lambda^2)}=\frac{1}{\lambda^2}\left\{\frac{2a\tau_{\rm s}}{1+4a^2\tau_{\rm s}^2}+\frac{\phi\sin(2a\phi)}{2\omega\tau_{\rm s}}\right\}\\
\end{equation}
where $\omega=\cos(2a\phi)-\cosh\left(\phi/\tau_{\rm s}\right)$.

We note that the first term inside the curly brackets depends only on $\lambda$, whereas the second term is also dependent on $\phi$. 

Lastly, the gradient of $\partial\Psi/\partial(\lambda^2)$ at $\phi$ is given by
\begin{align}
\label{eq:app5}
&\frac{\partial^2\Psi}{\partial(\lambda^2)\partial\phi}=\frac{1}{4\lambda^2}\times\nonumber\\
&\times\left\{\frac{4a\phi-2\cosh\left(\phi/\tau_{\rm s}\right) \left[2a \phi \cos(2a\phi)+\sin(2a\phi)\right]+\sin(4a\phi)}{\omega^2 \tau_{\rm s} }\right.+\nonumber\\ 
& \ \ \ \ \ \ \ +\left.\frac{2\phi\sin(2a\phi) \sinh\left(\phi/\tau_{\rm s}\right)}{\omega^2\tau_{\rm s}^2 }\right\}
\end{align}
We can assume values for $k$ and $a$, in order to get a handle of the magnitude of Eq.~\ref{eq:app3}, for different wavelengths. For the PA gradient, we have assumed $a=2\pi$, corresponding to a complete wrap of the PA per 0.5 rad (see Fig.~\ref{fig:app1}a). In addition, we assume $k=0.006$ rad m$^{-4}$, corresponding to $\tau_{\rm s}\approx 0.5$ rad at 100 MHz. The exponential functions with which we have convolved our profile are shown for the first phase bin and at different frequencies, in Fig.~\ref{fig:app1}b.

\begin{figure}
\centering
\includegraphics[height=\dimexpr \textheight - 10\baselineskip\relax,width=0.47\textwidth,keepaspectratio]{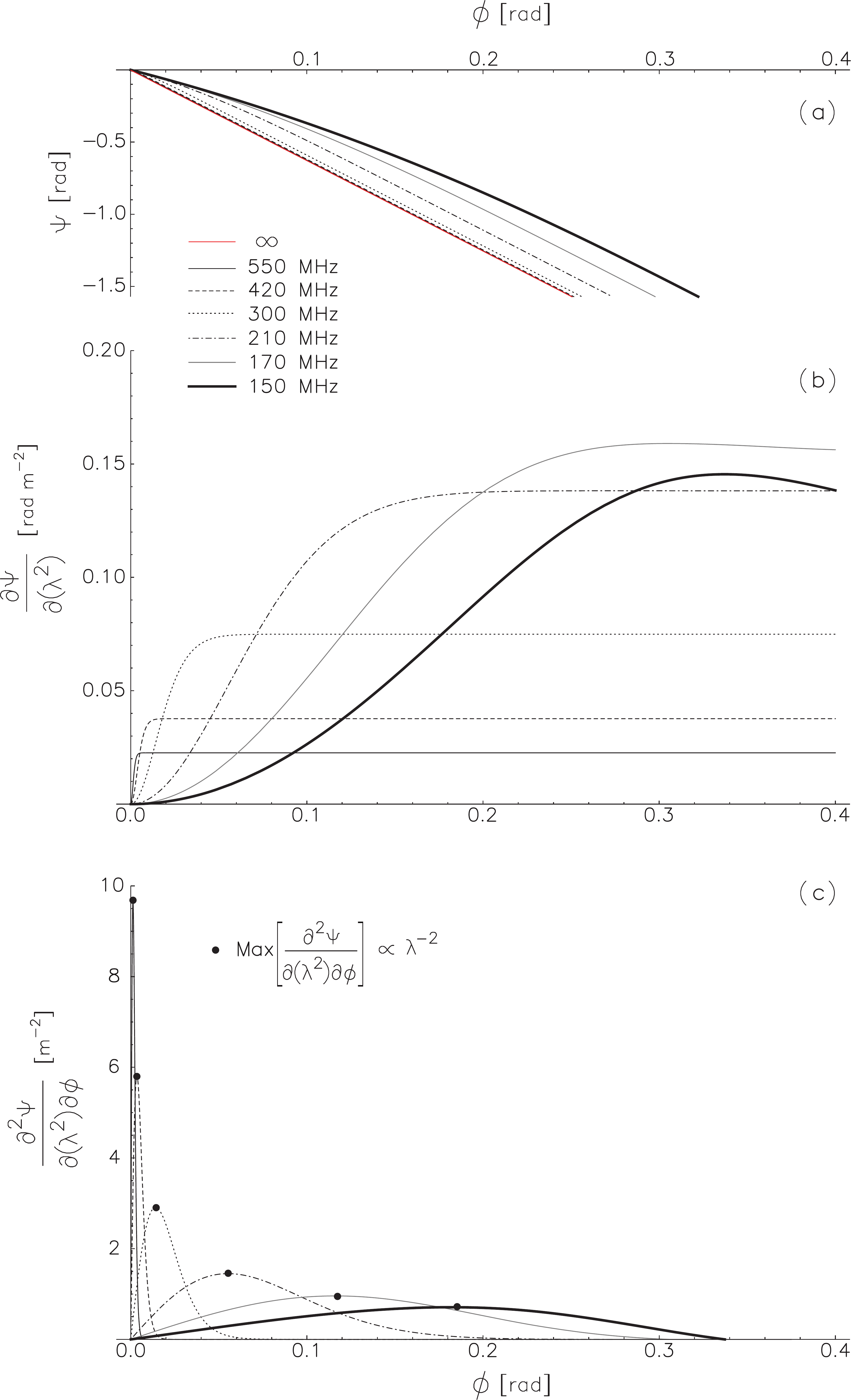}
\caption{(a) The PA profile at different frequencies, after convolution of the Stokes parameters with a one-sided exponential function, $f(\phi)={\rm e}^{-\phi/(k\lambda^4)}$, where $k=0.006$ rad m$^{-2}$. (b) The gradient of $\Psi(\lambda)$ at $\lambda^2$, as a function of pulse phase, for different frequencies. (c) The gradient of $\partial\Psi(\lambda)/\partial(\lambda^2)$ at pulse phase, $\phi$, as a function of pulse phase.}
\label{fig:app2}%
\end{figure}

Using the above values, we have plotted Eq.~\ref{eq:app5} between $\phi=0$ and 0.4 rad, for different values of $\lambda$ (see Fig.~\ref{fig:app2}c). Since we are interested in the maximum effect of scattering on the frequency evolution of the PA, we have marked the maxima of this function at the different frequencies. It can be shown that the ordinates of the maxima follow
\begin{equation}
{\rm Max}\left[\frac{\partial^2\Psi}{\partial(\lambda^2)\partial\phi}\right]\propto \frac{1}{\lambda^2}.
\end{equation}

\end{document}